\documentclass[aps,prb,floatfix,twocolumn,10pt,superscriptaddress]{revtex4-2}

\usepackage{graphicx}
\usepackage{float}
\usepackage{dcolumn}
\usepackage{bm}
\usepackage{soul}
\usepackage[dvipsnames]{xcolor}
\usepackage{amsmath}
\usepackage{amssymb}
\usepackage{amsmath}
\usepackage{float}
\usepackage{natmove}
\usepackage{multirow}
\usepackage{setspace}
\usepackage{array}
\usepackage{booktabs}
\usepackage{rotating}
\usepackage[colorlinks=true,linkcolor=blue,citecolor=blue,urlcolor=blue]{hyperref}
\usepackage{}
\usepackage{physics}
\usepackage{mathrsfs}
\usepackage{amssymb,amsfonts,dsfont}
\usepackage{bbold}
\usepackage{times} 
\usepackage{graphicx} 
\usepackage{accents} 
\usepackage{epic}
\usepackage{wrapfig}

\makeatletter

\begin{document}
\title{Binding and spontaneous condensation of excitons in narrow-gap carbon nanotubes}
\author{Giacomo Sesti}
\affiliation{FIM Department, Universit\`a degli Studi di Modena e Reggio Emilia, Via Campi 213a, 41125 Modena, Italy}
\affiliation{CNR-NANO, Via Campi 213a, 41125 Modena, Italy}
\author{Daniele Varsano}
\affiliation{CNR-NANO, Via Campi 213a, 41125 Modena, Italy}
\author{Elisa Molinari}
\affiliation{FIM Department, Universit\`a degli Studi di Modena e Reggio Emilia, Via Campi 213a, 41125 Modena, Italy}
\affiliation{CNR-NANO, Via Campi 213a, 41125 Modena, Italy}
\author{Massimo Rontani}
\affiliation{CNR-NANO, Via Campi 213a, 41125 Modena, Italy}

\begin{abstract}
    Ultraclean, undoped carbon nanotubes are observed to be always insulating, even when the gap predicted by band theory is zero: the residual band gap is then thought to have a many-body origin. Here we theoretically show that the correlated insulator is excitonic in {\it all stable} narrow-gap tubes irrespective of their size, thus extending our previous claim, limited to gapless (armchair) tubes
    [D.~Varsano, S.~Sorella, D.~Sangalli, M.~Barborini, S.~Corni, E.~Molinari, M.~Rontani, Nature Communications {\bf 8,} 1461 (2017)]. 
    We derive the scaling law of the exciton binding energy with the tube radius and chirality, and compute self-consistently the fundamental transport gap of the excitonic insulator, by 
    enhancing the two-band model with an accurate treatment of screening validated from first principles. Our findings point to the broader connection between the exciton length scale, dictated by structure, and the stability of the excitonic phase.       

\end{abstract}

\maketitle

Low-dimensional carbon allotropes, such as fullerenes \cite{Smalley1997,Andreoni2000}, nanotubes \cite{Dresselhaus1998,Charlier2007,McEuen2010,Deshpande2010,Laird2015}, and graphene \cite{Geim2007,CastroNeto2009,Abergel2010}, exhibit remarkable electronic and mechanical properties, which hold promise for novel physics and applications. Fascinatingly, they may host strongly-interacting electron phases 
that fall outside the traditional realm of oxides and narrow-band materials. Examples include Wigner crystal \cite{Deshpande2008,Secchi2009,Deshpande2010,Secchi2010,Pecker2013,Ilanit2019,Lotfizadeh2019,Ziani2021}, Luttinger liquid \cite{Giamarchi2004,Balents1997,Kane1997b,Egger1997,Krotov1997,Yoshioka1999,Bockrath1999,Postma2001,Deshpande2010}, unconventional superconductors \cite{Cao2018,Yankowitz2019,Hao2021,Park2021,Zhou2022}, and a plethora of correlated insulators. Very interesting phenomena are induced through handles that flatten the bands, including magnetic field \cite{Liu2017,Li2017,Liu2022,Li2024} and layer twisting \cite{Cao2018bis,Sharpe2019,Stepanov2020,Saito2020,Liu2020,Shen2020,He2021,Chen2021,Rickhaus2021}. 

A relevant open question now concerns those many-body phases that point to \textit{intrinsic} electronic correlations in carbon allotropes, i.e., in the absence of such external handles leading to flat bands. 
Their hallmark signature is the many-body gap that opens at the Dirac point and charge neutrality, which is now seen in suspended nanotubes \cite{Deshpande2009,Senger2018,island2018} and graphene bilayers stacked in Bernal arrangement \cite{Feldman2009,Freitag2012,Velasco2012,Bao2012,Geisenhof2021} (noticeably, no gap is observed in monolayer graphene).

The origin of this gap remains a subject of debate. An obvious scenario is the Mott insulator \cite{Deshpande2009,Cao2018bis}. Peierls \cite{Bohnen2004,Connetable2005,Chen2008,dumont2010peierls},  topological \cite{Efroni2017}, and other phases \cite{Voliovich2022,Kostyrko2024}  were also proposed for nanotubes, and various quantum Hall state flavors for bilayers \cite{Zhang2011}.
However, the large energy width of the valence bands, the long range of the unscreened Coulomb interaction, and
the widespread presence \cite{Maultzsch2005,Wang2005,Kravets2010,Ju2017} of bound electron--hole ($e$--$h$) pairs--excitons,---which have been observed even in metallic systems \cite{Wang2007}--- support the alternative paradigm of the `excitonic insulator' (EI) \cite{Khveshchenko2001,varsano2017carbon}. 

The EI is a permanent Bose-Einstein condensate of excitons, and its many-body gap is the remnant of the exciton binding energy, as initially hypothesized by W.~Kohn and others \cite{Kohn1967,Halperin1968}. In Ref.~\onlinecite{varsano2017carbon} we have shown theoretically that gapless (armchair) tubes have EI nature, as inferred by first-principles many-body perturbation theory and confirmed by accurate Quantum Monte Carlo simulations. The EI might also couple with a concurrent lattice distortion through electron-phonon interaction \cite{Hellgren2018,Okamoto2018,Barborini2022}. 

To address this question, here we focus on narrow-gap carbon nanotubes. By means of theory validated from the first principles, we demonstrate that all mechanically stable tubes \cite{elliott2004} are EIs. Our findings highlight the key role of the exciton length scale ruling the binding and also shed light onto other carbon-based systems.

\begin{figure}[b]
 \includegraphics[scale=0.26,trim={0.0cm 0.0cm 0.0cm 0.0cm},clip]{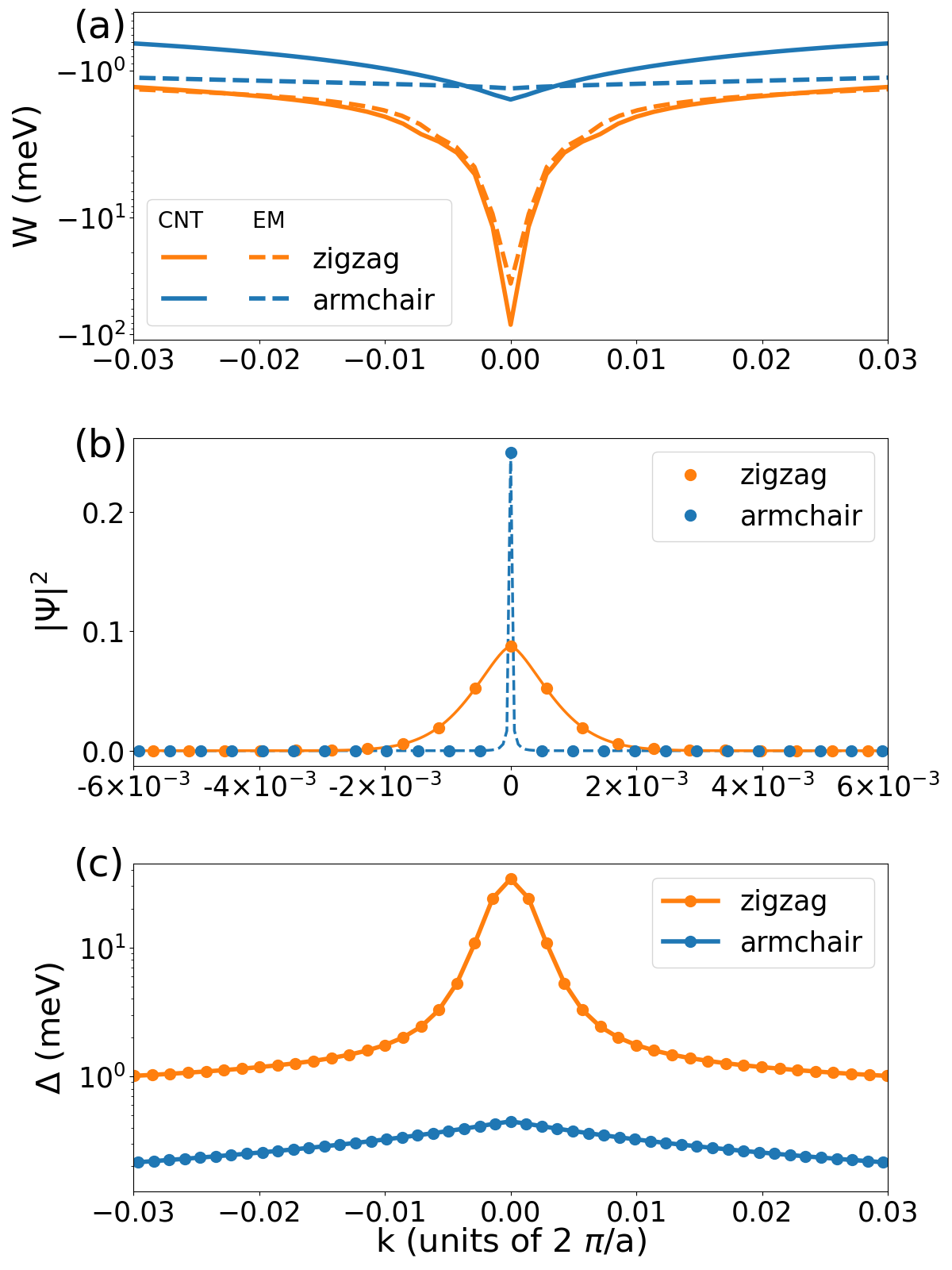} 
\caption{Attractive Coulomb force and exciton wave function. (a) Statically screened Coulomb interaction, $W(k)$, vs momentum parallel to the axis, $k$ (in reciprocal lattice units, with $a=$ 2.46 \AA). The dressed potential $W$ is computed within either an accurate model for the nanotube screening, validated from first principles and including local fields \cite{Sesti2022} (CNT, solid line), or the effective mass approximation (EM, dashed line).  
(b) Exciton wave function, $\left|\psi(k)\right|^2$. The axis origin is the Dirac point.
(c) Interband coherence of the excitonic insulator, $\left|\Delta(k)\right|$, for the (18,0) zigzag and (10,10) armchair tubes, which are respectively gapped and gapless. 
\label{f_wf}}
\end{figure}

According to band theory, two thirds of single-walled carbon nanotubes are semiconductors, 
with a (primary) gap of the order $\sim 1$ eV, always larger than the exciton binding energy, $E_B$ \cite{Ando1997,Spataru2004,Maultzsch2005}. Therefore, excitons cannot spontaneously form, and the tubes are simple band insulators.
The remaining one third of the tubes is potentially affected by spontaneous excitons generation, since the (secondary) gap, induced by curvature, is zero or takes a small value of the order of $\sim 0.1$ eV \cite{kane1997size,Ando1997,Charlier2007,Laird2015}. 

In his seminal theory of excitons, Ando ruled out the EI phase \cite{Ando1997}, based on an effective mass treatment of the screened $e$-$h$ Coulomb attraction, $W$. His argument is most easily seen in the reciprocal space, $k$. In the presence of a gap,
$W(k)$ is singular at long wave lengths, $W\sim \log(kR)$, as apparent from the orange curve of Fig.~\ref{f_wf}(a) (the vertical axis scale is logarithmic and $R$ is the tube radius). This turns into strong exciton binding, $E_B >0$. As the gap closes, the singularity is suppressed, $W$ is almost constant,  [blue dashed curve in Fig.~\ref{f_wf}(a)], and $E_B$ vanishes. This agrees with the elementary Thomas-Fermi picture of metallic screening, as there are now two Fermi points. However, we demonstrated~\cite{Sesti2022} that local field corrections, enhanced by tube topology, critically modify screening behavior, protecting the singularity of $W$ (solid blue curve): this results in the EI nature of gapless tubes~\cite{varsano2017carbon}.

This paper focuses on the fate of the nanotubes with a narrow curvature gap, and will show that they also result to be excitonic insulators. 
The electronic structure of nanotubes descends from graphene, with sub-bands quantized by rolling the honeycomb sheet into a cylinder \cite{ajiki1993electronic,Ando1997,Dresselhaus1998,Charlier2007,Laird2015}. In narrow-gap tubes, the folded bands feature gapped Dirac cones at the corners of the Brillouin zone K and K$'$. The envelopes of the topmost valence ($v$) and lowest conduction ($c$) sub-bands, $ \boldsymbol{F}^{\tau}_{\alpha k }$, satisfy the Dirac equation 
\begin{align}
\label{eq.Dirac}
\gamma \,\tau \left[ \boldsymbol{\sigma}_x  k_c + \boldsymbol{\sigma}_y k \right]
\boldsymbol{F}^{\tau}_{\alpha k  } &= \varepsilon_{\alpha}(k)\, \boldsymbol{F}^{\tau}_{\alpha k  },
\end{align}
where $\alpha=c,v$ labels the band,
$\tau=1$ ($\tau=-1$) for K (K$'$),  $\gamma$ is the graphene band parameter, $\boldsymbol{\sigma}_x$
and $\boldsymbol{\sigma}_y$
are Pauli matrices acting on the two-component spinor $ \boldsymbol{F}$ in the sublattice space, and $k$ is the wavevector along the tube axis, reckoned from the Dirac point $\tau$. 
The electrons exhibit the Dirac dispersion $\varepsilon_{\alpha}(k)=\pm \gamma \left[k_c^2+k^2\right]^{1/2}$, with the sign $+$ ($-$) corresponding to $\alpha=c$ ($\alpha=v$). The bare curvature gap, $\varepsilon_g^{\text{bare}}=2 \gamma k_{c}$, is obtained for $k=0$. This gap depends \cite{kane1997size,Laird2015} on both the radius of the nanotube, $R$, and the chiral angle, $\theta$, as
\begin{eqnarray} 
\label{eq.kc}
k_{c} = (0.625 \textrm{ eV} \cdot \textrm{\AA}) \cos(3 \theta) / \gamma R^2. 
\end{eqnarray}

The chiral angle encodes the way the graphene sheet is rolled up to make a cylinder, ranging from 
$\theta=0$ (zigzag tube) to
$\theta=\pi/6$ (armchair, gapless). Hereafter we consider continuous values of $R$ and $\theta$, which include those discrete values of the narrow-gap physical tubes established from the (\textit{n}, \textit{m}) indices \cite{ajiki1993electronic,Ando1997,Dresselhaus1998,Charlier2007,Laird2015}.

We compute the lowest-energy excitons, of momentum zero and spin one, by solving the Bethe-Salpeter equation (BSE) \cite{Ando2006,varsano2017carbon},
\begin{eqnarray}
\label{eq.bse}
&&\left[ 2\gamma \sqrt{k_c^2+k^2} 
+ \Sigma(k)
\right] \psi_{\tau}(k)  \nonumber \\ 
&&-\quad \frac{1}{A} \sum_{q} (\boldsymbol{F}^{\tau \dagger}_{v k} \boldsymbol{F}^{\tau}_{v k+q})   (\boldsymbol{F}^{\tau \dagger }_{c k+q} \boldsymbol{F}^{\tau}_{c k}) W(q)  \;\psi_{\tau}(k+q) \nonumber  \\ && \qquad - \quad  \frac{1}{A}   \sum_{\tau' \neq \tau } \sum_{q} W^{\tau\tau'} \psi_{\tau'\!}(k+q) \;\; =\;\;  \mathcal{E}_u\; \psi_{\tau}(k).
\end{eqnarray}
The excitation energy
and wave function of the exciton are $\mathcal{E}_u$ and $\psi_{\tau}$, respectively, and $A$ is the tube length.
To achieve predictive accuracy for those large-$R$ tubes that cannot be handled from first principles \cite{varsano2017carbon}, we improve the model two-band BSE \cite{Ando1997,Ando2006} in two respects. First, the screened long-range interaction $W$ is evaluated like the solid blue curve in Fig.~\ref{f_wf}(a), i.e., including local field effects, assessed through first-principles validation \cite{Sesti2022}. Second, a correction, $\Sigma(k)$, adds to the energy of the free $e$-$h$ pair excited through transfer of an electron of momentum $k$ from $v$ to $c$ band. This self energy is due to the Coulomb dressing of quasiparticles, according to:  
\begin{multline}
\Sigma(k)
 = \frac{1}{3 A} \sum_q (| \boldsymbol{F}^{\tau \dagger}_{v k} \boldsymbol{F}^{\tau}_{v k+q}|^2-| \boldsymbol{F}^{\tau \dagger }_{c k} \boldsymbol{F}^{\tau}_{v k+q}  |^2 ) W(q),
\end{multline}
as validated by the first-principles GW study of Ref.~\onlinecite{island2018} (see Appendix \ref{sec:Self}).
Since $\Sigma$ may be even larger than the magnitude of $\varepsilon_g^{\text{bare}}$, it effectively counteracts the tendency of excitons to spontaneously form. 
Finally, we include the intervalley short-range part of the Coulomb potential \cite{Ando2006}, $W^{\tau\tau'}$, which weakly depends on $k$, like in Ref.~\citenum{varsano2017carbon}.

\begin{figure}[t]
\includegraphics[scale=0.245,trim={1.0cm 0.2cm 4.0cm 2.5cm},clip]{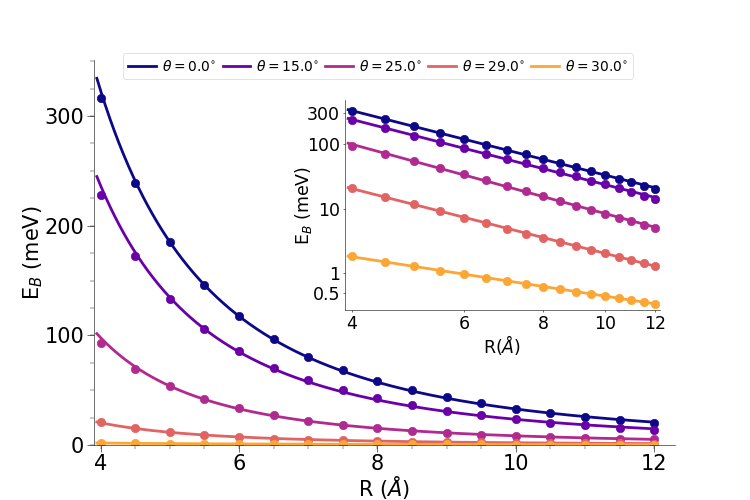}
\caption{Exciton binding energy, $E_B$, vs radius, $R$, for narrow-gap carbon tubes. The chiral angle, $\theta$, varies between $\theta=0$ (zigzag tubes) and $\theta=\pi/6$ (armchair). Armchair tubes are gapless due to symmetry.  The log-log plot (inset) shows that $E_B\sim R^s$, with $s=-2.54$ and $-1.53$ $(\pm 0.07)$  for chiral and armchair tubes, respectively. The calculations were done including only the regions of the Brillouin zone around the two Dirac points with an extension 0.1 $\times 2 \;\pi/a$, each region discretized with 41 $k$ points. The results have proved equivalent to selected calculations on the full Brillouin zone with 801 $k$ points. \label{f_binding}} 
\end{figure}

The exciton binding energy, $E_B= \varepsilon_{g} -\mathcal{E}_u$, is obtained
after solving the BSE and renormalizing the energy band gap,
$\varepsilon_g=\varepsilon_g^{\text{bare}}+\Sigma(0)$.
Figure \ref{f_binding} shows $E_B$ vs radius for selected angles $\theta$, in the range of $R$ that corresponds to mechanically stable \cite{elliott2004} tubes relevant to experiments \cite{Deshpande2009,Senger2018}.
The results for gapless tubes ($\theta = \pi/6$, yellow points) provide a first insight into the excitonic instability. According to the textbook treatment of excitons in bulk semiconductors, the binding energy, $E_B^{\textrm{bulk}} \sim m/\kappa_r^2 $, is a combination of the effective mass, $m$, and relative dielectric constant, $\kappa_r$, and is inversely proportional to the characteristic length scale of the exciton (Bohr radius, $a_B$), $E_B^{\textrm{bulk}} \sim 1/(a_B^{\textrm{bulk}} \kappa_r)$. Since electrons in gapless tubes are massless, one expects $E_B^{\textrm{bulk}}=0$ and
$a_B^{\textrm{bulk}}=\infty$. However, $E_B$ does not vanish \cite{varsano2017carbon} but exhibits a neat dependence on $R$, $E_B\sim 1/R^{1.53}$, as apparent from the Arrhenius plot in the inset. 

To understand the scaling law of this ``Dirac'' exciton, one may at first neglect self-energy and  screening effects in \eqref{eq.bse}. Then, the BSE is scale invariant with respect to $R$, that is
$\mathcal{E}_u = \gamma E_0/R$, with $E_0$ being a pure number derived from the dimensionless BSE (see Supplementary Material of Ref.~\onlinecite{varsano2017carbon}). This number is small, $E_0 \sim 10^{-3} $, and only depends on the fine structure constant of graphene, $\alpha= e^2/\gamma$. In summary, for unscreened interaction, $E_B\sim 1/R$, which is not too far from the scaling in the presence of screening [the discrepancy arises from the intricate polarization behavior, cf.~Eq.~(30) of Ref.~\citenum{Sesti2022}].
The exciton length scale, $a_B$, may be assessed from Heisenberg's uncertainty principle, $a_B \sim 1/(\Delta k)$, as the indetermination in $k$ space descends from binding, $\Delta k \sim E_B / \gamma$. The resulting Bohr radius, $a_B \sim (1/E_0) R^{1.53} $, points to the extreme localization (delocalization) of $\left|\psi\right|^2$ in momentum (real) space, as apparent from the blue curve of Fig.~\ref{f_wf}(b). 

\begin{figure}[t]
\includegraphics[scale=0.225,trim={0.0cm 0.0cm 0.5cm 0.0cm},clip]{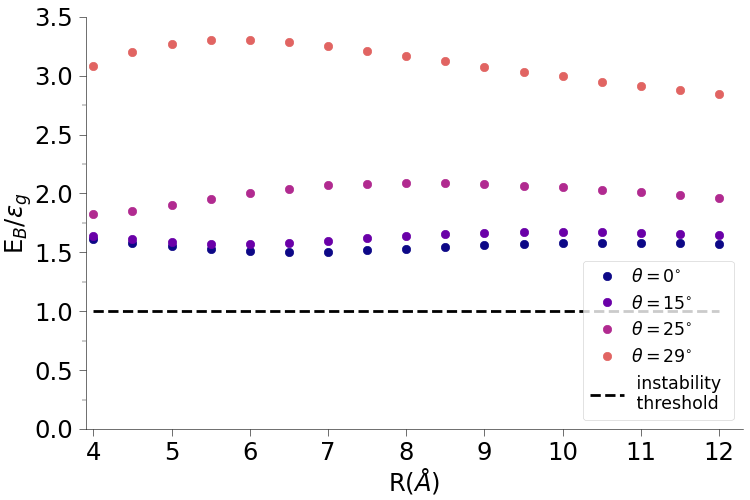}
\caption{
Ratio of exciton binding energy, $E_B$, to quasiparticle gap, $\varepsilon_{g}$, as a function of radius, $R$, for representative chiralities. The dashed line indicates the threshold for the excitonic instability.}
\label{f_ratio}
\end{figure}
  
The finiteness of $E_B$ implies that gapless tubes are unstable against the
spontaneous condensation of excitons. The many-body gap of the resulting EI phase has the same dependence on $R$ as $E_B$. Also, the gap of the Mott insulator depends on $R$, since it equally descends from electron-electron interaction.  However, the scaling exponent, $-1/(1-g)$, is in principle different ($g$ is the Luttinger liquid parameter) \cite{Kane1997b,varsano2017carbon}. Whereas for the EI the dependence on $R$ descends from the screened, long-range term, $W$, for the Mott case the origin is the short-range, Hubbard-like part.
 
Figure \ref{f_binding} shows that, by lowering the chirality angle $\theta$ moving away from the upper bound $\theta=\pi/6$, $E_B$ significantly increases, by even two orders of magnitude as $\theta$ approaches the zigzag lower bound ($\theta=0$, blue curve). The scaling exponent changes as well, E$_B \sim R^{-2.54 \pm 0.07}$, and hardly depends on $\theta$. This points to qualitatively different physics. Since excitons are now massive, it is tempting to use the familiar bulk formula for $E_B^{\textrm{bulk}}$. One issue is how to map band curvature and screening into parameters $m$ and $\kappa_r$, respectively. 

A key observation is that the transverse momentum $k_c$, which controls the curvature gap magnitude through \eqref{eq.kc}, fixes the crossover from bare ($k<k_c$) to screened interaction ($k>k_c$) as one swipes $W(k)$ through the momentum space \cite{Sesti2022}. This can be directly checked from the orange curve of Fig.~\ref{f_wf}(a) for the (18,0) zigzag tube, with $k_{c}\simeq5\times10^{-3} \ 2\pi/a$. In the range $0<k<k_c$, the screened long-range interaction $W(k)$ is linear (on the log scale), corresponding to the bare Coulomb interaction in the tube \cite{Ando1997}, whereas for $k>k_c$ the slope of the $e$-$h$ interaction changes, as the force senses the electron polarization. Therefore, the major contribution to exciton binding comes from the scattering of unscreened $e$-$h$ pairs with $k<k_c$. Also note that, in this same range of $k$, we may safely approximate the kinetic energy of $e$-$h$ pairs through Taylor expansion, 
which provides the effective mass $m\sim \varepsilon_{g}/2\gamma^2 $. In summary, $k_c$ simultaneously controls band curvature 
and screening, hence  
the bulk result for the binding energy applies
to massive Dirac excitons as well, in the form $E_B\sim \varepsilon_{g}$ and $a_B \sim 1/\varepsilon_{g}$.
 
This expectation is confirmed by Fig.~\ref{f_ratio}, which shows the ratio $E_B /\varepsilon_{g}$  vs $R$ for fixed chiral angle $\theta$. Such ratio is almost constant with respect to $R$, weakly oscillating between 1.5 and 2 for all chiral angles $\theta$,  apart from those tubes that are close to the gapless limit ($\theta=29^\circ$, red points). Importantly, the ratio is always larger than one, hence excitons
spontaneously form in {\it all} stable narrow-gap tubes, stabilizing the EI phase with respect to the pristine semiconductor.

\begin{figure}[t!]
\centering
\begin{tabular}{cc}
\large{(a)}   &  \\
 &  \hspace{-1cm} \includegraphics[scale=0.24,trim={0.0cm 0.0cm 0.0cm 0.0cm},clip]{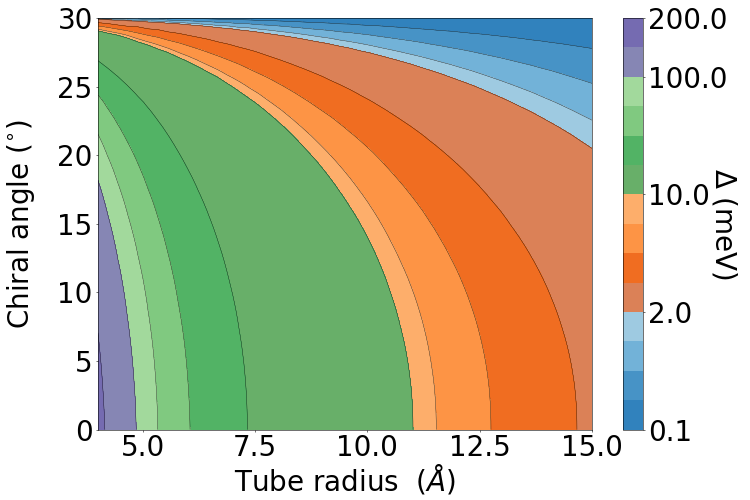}\\
\large{(b)}   &   \\
 &  \hspace{-1cm}  \includegraphics[scale=0.24,trim={0.0cm 0.0cm 0.0cm 0.0cm},clip]{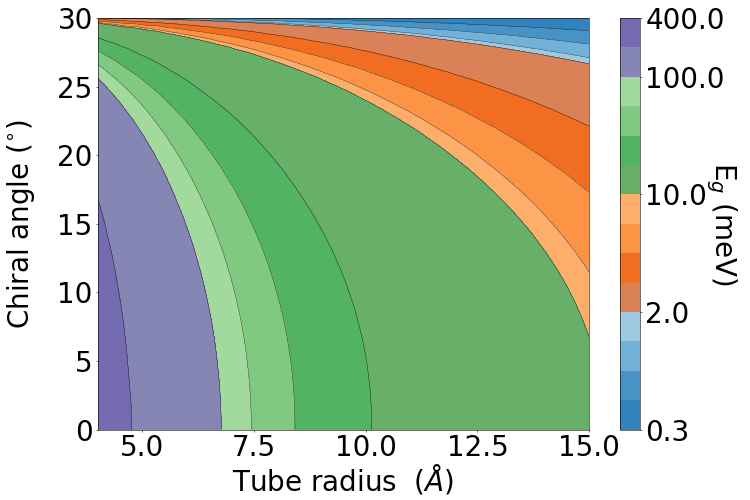}
\end{tabular}
\caption{ Contour map of (a) interband coherence at Dirac point, $\Delta(k=0)$ (b) fundamental gap of the excitonic insulator, $E_{g}$, in the $(R,\theta)$ space. \label{f_contour}}
\end{figure}

In the EI phase the fundamental transport gap, $E_g$, opens according to $E_{g} = \left[\epsilon_{g}^2 + 4 |\Delta(k=0)|^2\right]^{1/2} $, due to the additional many-body contribution, the 
interband coherence, $\Delta(k)$, evaluated at the Dirac Point $k=0$.
The coherence is roughly proportional to the probability density for an $e$-$h$ pair of momentum $k$ populating the condensate. Whereas $\psi$ is the normalized solution of the BSE \eqref{eq.bse}, $\Delta$ is evaluated self-consistently, by solving a pseudo-BSE (gap equation) 
within a BCS-like variational framework, and is not normalized (see Appendix \ref{sec:Exc-Gap}). The comparison of Figs.~\ref{f_wf}(b) and (c)
shows that $\Delta$ is more rounded than $\psi$ around $k=0$, because Pauli blockade sets an upper bound on the number of condensing $e$-$h$ pairs for given $k$ \cite{Keldysh1995}.   

The contour map \ref{f_contour} shows the computed values of
$\left|\Delta(k=0)\right|$ (panel a) and $E_{g}$ (panel b) in the space whose axes are $R$ and $\theta$. 
The behavior of $\left|\Delta\right|$ is very similar to the one found for $E_{B}$, in particular the interband coherence scales with $R$ almost like $E_B$, attaining maximum value for small radii and zigzag chirality. We find $\left|\Delta\right|\sim R^{-1.51\pm0.1}$ for armchair tubes while
$\left|\Delta\right|\sim R^{-2.75\pm0.1}$ for other chiralities.
On the other hand, the gap $E_{g}$ is less sensitive to chirality than $\left|\Delta\right|$. The dependence of $E_{g}$ on $\theta$ becomes mild for larger radii, the gap being nearly constant over a wide range of $\theta$ for $R > 1$ nm.  Remarkably, in very small nanotubes ($R < 0.5$ nm) $E_{g}$ can be quite large, even above 400 meV.  

The prediction of Fig.~\ref{f_contour} may be used to pinpoint experimentally the EI phase, by means of performing a low-temperature transport measurement of $E_{g}$, like in the seminal work \cite{Deshpande2009} by Deshpande {\it et al.}, supplemented by the knowledge of tube size and chirality. Intriguingly, it was recently suggested \cite{Senger2018} that there is no correlation between $\theta$ and $E_{g}$. A direct comparison with our theory is difficult, since the electric characteristics of the tubes, used to infer $E_{g}$ through a thermal activation model, were collected at room temperature. In general, the significant discrepancies \cite{Deshpande2009,Senger2018} concerning the measured size of the many-body gap call for next-generation experiments of superior precision. 

Ideally, the transition to the EI phase is a purely electronic reconstruction of the ground state, accompanied by a disproportion of the charge between the A and B honeycomb sublattices (in the absence of significant spin effects \cite{varsano2017carbon}). The exciton polarization built in the condensate displaces the charge, whose variation magnitude and sign are related to the modulus and phase of $\Delta$, respectively. Then, a lattice distortion, driven by the electron-phonon interaction, may couple with the A-B charge displacement, giving rise to a hybrid exciton Peierls scenario \cite{Hellgren2018,Barborini2022}. This coupling makes exciton condensation hard to assess, as it requires separating between exciton and phonon contributions to the gap or the A-B symmetry breaking.
Importantly, the lattice contribution is expected only for the smallest tubes, 
since it scales like $R^{-3}$ \cite{Chen2008} and hence becomes negligible for $R > 1$ nm
\cite{Bohnen2004,Connetable2005,Chen2008,dumont2010peierls,Hellgren2018,Okamoto2018}. Therefore, for most part of the experimentally relevant tubes, in the range 0.8 $< R < $ 1.5 nm \cite{Deshpande2009,Senger2018}, the predicted value for $\Delta$ fully exhausts the size of $E_{g}$. Alternative measurable~\cite{Khivrich2019} fingerprints of the EI include the collective oscillations of the amplitude and phase of $\Delta(k)$, namely, a rich low-energy exciton-like spectrum, and a peculiar acoustic mode, the `exciton sound'. The velocity of the latter is of the order of $\gamma/\hbar$, much higher than that of the ordinary sound.

\begin{figure}
    \centering
    \includegraphics[scale=0.24,trim={0.5cm 0.2cm 0.0cm 0cm},clip]{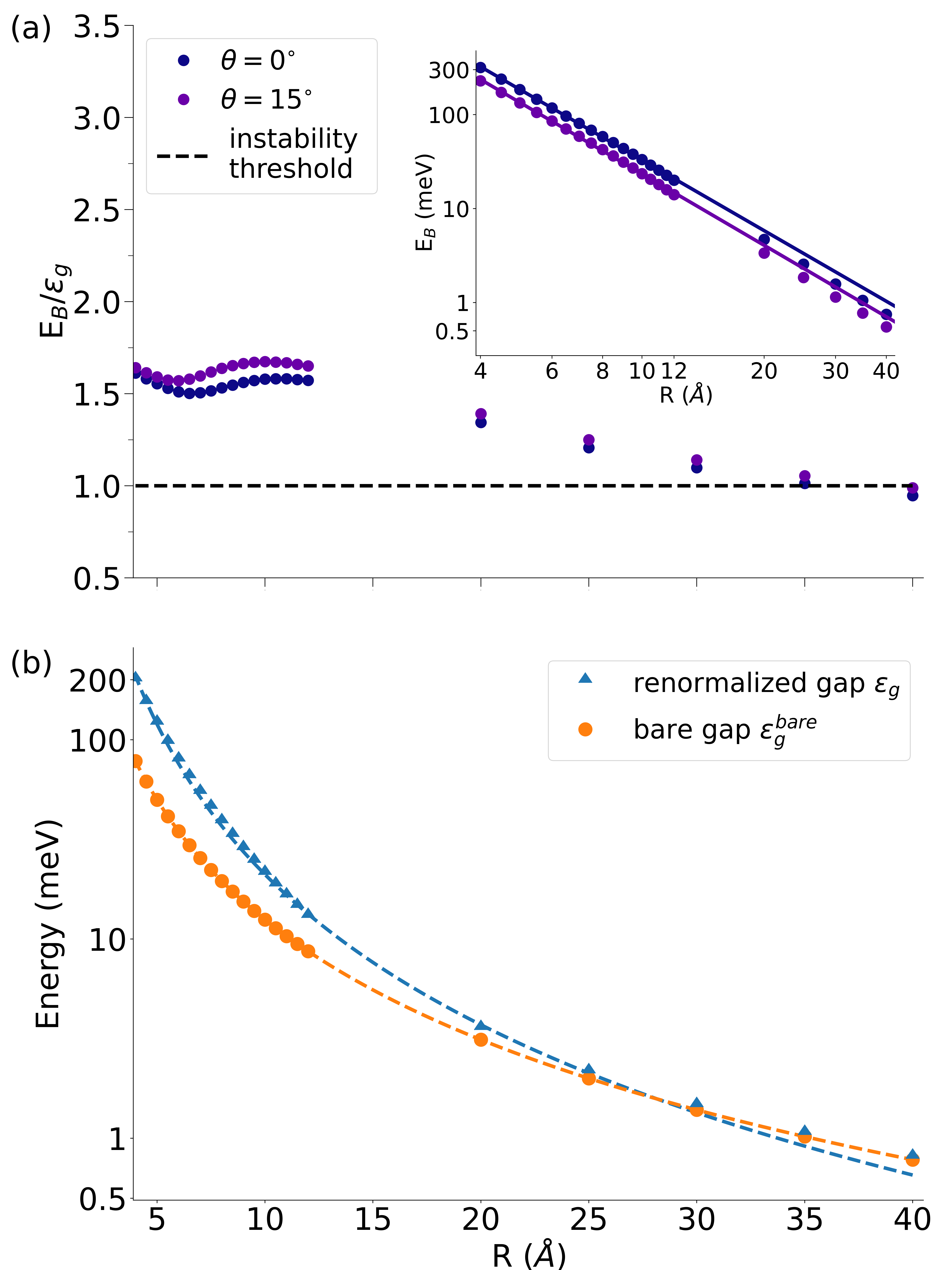}
    \caption{Scaling properties of tubes of very large radius. (a) Ratio of the exciton binding energy, $E_B$, to quasiparticle gap, $\varepsilon_{g}$, as a function of radius, $R$, for chiral angle $\theta=0$ (dark dots) and $\theta=15^{\circ}$ (light dots). The dashed line indicates the threshold for excitonic instability. Inset: Scaling of $E_B$ (log-log plot). The lines correspond to the scaling exponent $s=-2.54$, with $E_B\sim R^{s}$. (b) Comparison in semilog scale of the renormalized, $\varepsilon_g$, and bare curvature gap, $\varepsilon_g^{\text{bare}}$, for zigzag tubes ($\theta=0$). The blue and orange dashed lines correspond to the scaling $R^{-2.5}$ and $R^{-2}$, respectively ($\varepsilon_g^{\text{bare}}\sim 1/R^2$).  
    \label{fig:crit_ratio}}
\end{figure}

Our findings for narrow-gap tubes provide a fresh clue to the question of exciton condensation in graphene \cite{Khveshchenko2001,CastroNeto2009b,Drut2009,Gamayun2009,Sabio2010,Gronqvist2012}, a semiconductor whose gap, if present, is exceedingly small with respect to experimental accuracy. We saw above that condensing excitons exhibit
a characteristic length scale, $a_B$, which is related to either $R$ (armchair tubes) or a combination of $R$ and $\theta$ (all other tubes), i.e., to structural parameters of the system.  
There is no such a length scale for monolayer graphene, at least within a low-energy effective model. As we show below, the graphene behavior is recovered in the $R\rightarrow \infty$ limit for the tube, as Coulomb interaction remains long-ranged and $a_B$ becomes meaningless. 

We investigate numerically the limit $R \gg a$, with $a$ being graphene lattice constant, by means of including in our study also those tubes that are structurally unstable due to their excessive size (the stability threshold is around $R \approx $  20  \AA \, \cite{elliott2004}). We take as an upper bound of our investigation $R=$ 40 \AA, as for larger radii the numerical results become noisy. For gapless tubes ($\theta=\pi/6)$, $E_B$ scales with $R$ with the same power of Fig.~\ref{f_binding},  hence 
the binding vanishes in the monolayer limit and no instability occurs. For other tubes, the renormalized curvature gap  significantly slows with $R$ around 30 \AA,
the scaling changing from $\varepsilon_g\sim R^{-2.5}$ [blued dashed line in Fig.~\ref{fig:crit_ratio}(b)]  to $R^{-2}$ (orange dashed line). This is due to the vanishing of the self-energy with $\varepsilon_g^{\text{bare}}$, since in the gapless limit there is no correction at Dirac point, $\Sigma(k=0)\rightarrow 0$ \cite{CastroNeto2009}. Therefore, above the crossover $R\approx$ 30 \AA, renormalized and bare gaps overlap, $\varepsilon_g \approx \varepsilon_g^{\text{bare}}$. On the other hand, the scaling of $E_B$ remains almost unchanged [inset of Fig.~\ref{fig:crit_ratio}(a)]. The consequence is that the ratio $E_B/\varepsilon_g$ approaches the instability threshold, one, around $R\approx $ 40 \AA, as shown in Fig.~\ref{fig:crit_ratio}(a) for both $\theta=0$ and $\theta=15^{\circ}$. In summary, in the monolayer limit no instability occurs, irrespective of the chirality angle $\theta$.  

Strong-coupling theories for graphene associate exciton formation with the spatial collapse of bound $e$-$h$ complexes \cite{Gamayun2009,Sabio2010,Gronqvist2012}. This is analogous to the `fall on the center' problem for heavy atoms in quantum electrodynamics (QED) \cite{Berestetskii1982}. To deal with it, one must regularize Coulomb interaction at a distance shorter than the lattice constant (of the order of the nuclear radius in the original QED problem). Then, the exciton length scale is just the imposed short-range cutoff. This hints at ultraviolet-range phenomena that are out of the reach of envelope-function  approaches, and certainly go beyond the familiar understanding of Wannier excitons in semiconductors. On the contrary, bi- and few-layer graphene provides an obvious exciton length scale, the interlayer spacing, which makes exciton binding natural, and condensation a possibility \cite{Min2008b,Min2008,Zhang2008,Kharitonov2008,Nandkishore2010}.        

In summary, we investigated the whole landscape of isolated zero- and narrow-gap nanotubes, as prototype carbon systems where long-range electronic interactions rule the many-body ground state. Based on an accurate first-principles description of screening, we analyzed the possibility of an electronic instability due to the spontaneous condensation of excitons; investigated the resulting EI 
phase by a self-consistent mean-field theory formulated in terms of interband coherence; established the corresponding phase diagram.
As an outcome, we found the scaling law of the exciton binding energy for tubes of any size and chirality, leading to the conclusion that the long-sought excitonic insulator is the true ground state of nominally metallic, intrinsic tubes. This paves the way to further experimental tests.

\section*{Acknowledgements}
We are grateful to Vikram Deshpande, Assaf Hamo, Shahal Ilani, Ethan Minot, Catalin Spataru, and Gary Steel for inspiring discussions along the years. This work was partly supported by: 
the MaX -- MAterials design at the eXascale -- European Centre of Excellence, co-funded by the European High Performance Computing joint Undertaking (JU) and participating countries within the program HORIZON-EUROHPC-JU-2021-COE-1 (Grant No. 101093374); ICSC -- Centro Nazionale di Ricerca
in High Performance Computing, Big Data and Quantum Computing -- funded by the European Union through the Italian Ministry of University and Research under PNRR M4C2I1.4 (Grant No.~CN00000013).

We acknowledge ISCRA and ICSC for awarding this project access to the LEONARDO supercomputer, owned by the EuroHPC Joint Undertaking, hosted by CINECA (Italy).

\bibliography{./biblio/biblio-art}

\begin{thebibliography}{94}%
\makeatletter
\providecommand \@ifxundefined [1]{%
 \@ifx{#1\undefined}
}%
\providecommand \@ifnum [1]{%
 \ifnum #1\expandafter \@firstoftwo
 \else \expandafter \@secondoftwo
 \fi
}%
\providecommand \@ifx [1]{%
 \ifx #1\expandafter \@firstoftwo
 \else \expandafter \@secondoftwo
 \fi
}%
\providecommand \natexlab [1]{#1}%
\providecommand \enquote  [1]{``#1''}%
\providecommand \bibnamefont  [1]{#1}%
\providecommand \bibfnamefont [1]{#1}%
\providecommand \citenamefont [1]{#1}%
\providecommand \href@noop [0]{\@secondoftwo}%
\providecommand \href [0]{\begingroup \@sanitize@url \@href}%
\providecommand \@href[1]{\@@startlink{#1}\@@href}%
\providecommand \@@href[1]{\endgroup#1\@@endlink}%
\providecommand \@sanitize@url [0]{\catcode `\\12\catcode `\$12\catcode
  `\&12\catcode `\#12\catcode `\^12\catcode `\_12\catcode `\%12\relax}%
\providecommand \@@startlink[1]{}%
\providecommand \@@endlink[0]{}%
\providecommand \url  [0]{\begingroup\@sanitize@url \@url }%
\providecommand \@url [1]{\endgroup\@href {#1}{\urlprefix }}%
\providecommand \urlprefix  [0]{URL }%
\providecommand \Eprint [0]{\href }%
\providecommand \doibase [0]{https://doi.org/}%
\providecommand \selectlanguage [0]{\@gobble}%
\providecommand \bibinfo  [0]{\@secondoftwo}%
\providecommand \bibfield  [0]{\@secondoftwo}%
\providecommand \translation [1]{[#1]}%
\providecommand \BibitemOpen [0]{}%
\providecommand \bibitemStop [0]{}%
\providecommand \bibitemNoStop [0]{.\EOS\space}%
\providecommand \EOS [0]{\spacefactor3000\relax}%
\providecommand \BibitemShut  [1]{\csname bibitem#1\endcsname}%
\let\auto@bib@innerbib\@empty
\bibitem [{\citenamefont {Smalley}(1997)}]{Smalley1997}%
  \BibitemOpen
  \bibfield  {author} {\bibinfo {author} {\bibfnamefont {R.~E.}\ \bibnamefont
  {Smalley}},\ }\bibfield  {title} {\bibinfo {title} {Discovering the
  fullerenes},\ }\href {https://doi.org/10.1103/RevModPhys.69.723} {\bibfield
  {journal} {\bibinfo  {journal} {Rev. Mod. Phys.}\ }\textbf {\bibinfo {volume}
  {69}},\ \bibinfo {pages} {723} (\bibinfo {year} {1997})}\BibitemShut
  {NoStop}%
\bibitem [{\citenamefont {Andreoni}(2000)}]{Andreoni2000}%
  \BibitemOpen
  \bibfield  {author} {\bibinfo {author} {\bibfnamefont {W.}~\bibnamefont
  {Andreoni}},\ }\href@noop {} {\emph {\bibinfo {title} {The physics of
  fullerene-based and fullerene-related materials}}}\ (\bibinfo  {publisher}
  {Springer},\ \bibinfo {address} {Berlin},\ \bibinfo {year}
  {2000})\BibitemShut {NoStop}%
\bibitem [{\citenamefont {Saito}\ \emph {et~al.}(1998)\citenamefont {Saito},
  \citenamefont {Dresselhaus},\ and\ \citenamefont
  {Dresselhaus}}]{Dresselhaus1998}%
  \BibitemOpen
  \bibfield  {author} {\bibinfo {author} {\bibfnamefont {R.}~\bibnamefont
  {Saito}}, \bibinfo {author} {\bibfnamefont {G.}~\bibnamefont {Dresselhaus}},\
  and\ \bibinfo {author} {\bibfnamefont {M.~S.}\ \bibnamefont {Dresselhaus}},\
  }\href@noop {} {\emph {\bibinfo {title} {Physical Properties of Carbon
  Nanotubes}}}\ (\bibinfo  {publisher} {Imperial College Press},\ \bibinfo
  {address} {London},\ \bibinfo {year} {1998})\BibitemShut {NoStop}%
\bibitem [{\citenamefont {Charlier}\ \emph {et~al.}(2007)\citenamefont
  {Charlier}, \citenamefont {Blase},\ and\ \citenamefont
  {Roche}}]{Charlier2007}%
  \BibitemOpen
  \bibfield  {author} {\bibinfo {author} {\bibfnamefont {J.-C.}\ \bibnamefont
  {Charlier}}, \bibinfo {author} {\bibfnamefont {X.}~\bibnamefont {Blase}},\
  and\ \bibinfo {author} {\bibfnamefont {S.}~\bibnamefont {Roche}},\ }\bibfield
   {title} {\bibinfo {title} {Electronic and transport properties of
  nanotubes},\ }\href {https://doi.org/10.1103/RevModPhys.79.677} {\bibfield
  {journal} {\bibinfo  {journal} {Rev. Mod. Phys.}\ }\textbf {\bibinfo {volume}
  {79}},\ \bibinfo {pages} {677} (\bibinfo {year} {2007})}\BibitemShut
  {NoStop}%
\bibitem [{\citenamefont {Ilani}\ and\ \citenamefont
  {McEuen}(2010)}]{McEuen2010}%
  \BibitemOpen
  \bibfield  {author} {\bibinfo {author} {\bibfnamefont {S.}~\bibnamefont
  {Ilani}}\ and\ \bibinfo {author} {\bibfnamefont {P.~L.}\ \bibnamefont
  {McEuen}},\ }\bibfield  {title} {\bibinfo {title} {Electron transport in
  carbon nanotubes},\ }\href
  {https://doi.org/https://doi.org/10.1146/annurev-conmatphys-070909-103928}
  {\bibfield  {journal} {\bibinfo  {journal} {Ann. Rev. of Cond. Mat. Phys.}\
  }\textbf {\bibinfo {volume} {1}},\ \bibinfo {pages} {1} (\bibinfo {year}
  {2010})}\BibitemShut {NoStop}%
\bibitem [{\citenamefont {Deshpande}\ \emph {et~al.}(2010)\citenamefont
  {Deshpande}, \citenamefont {Bockrath}, \citenamefont {Glazman},\ and\
  \citenamefont {Yacoby}}]{Deshpande2010}%
  \BibitemOpen
  \bibfield  {author} {\bibinfo {author} {\bibfnamefont {V.~V.}\ \bibnamefont
  {Deshpande}}, \bibinfo {author} {\bibfnamefont {M.}~\bibnamefont {Bockrath}},
  \bibinfo {author} {\bibfnamefont {L.~I.}\ \bibnamefont {Glazman}},\ and\
  \bibinfo {author} {\bibfnamefont {A.}~\bibnamefont {Yacoby}},\ }\bibfield
  {title} {\bibinfo {title} {Electron liquids and solids in one dimension},\
  }\href {https://doi.org/https://doi.org/10.1038/nature08918} {\bibfield
  {journal} {\bibinfo  {journal} {Nature}\ }\textbf {\bibinfo {volume} {464}},\
  \bibinfo {pages} {209} (\bibinfo {year} {2010})}\BibitemShut {NoStop}%
\bibitem [{\citenamefont {Laird}\ \emph {et~al.}(2015)\citenamefont {Laird},
  \citenamefont {Kuemmeth}, \citenamefont {Steele}, \citenamefont
  {Grove-Rasmussen}, \citenamefont {Nyg\aa{}rd}, \citenamefont {Flensberg},\
  and\ \citenamefont {Kouwenhoven}}]{Laird2015}%
  \BibitemOpen
  \bibfield  {author} {\bibinfo {author} {\bibfnamefont {E.~A.}\ \bibnamefont
  {Laird}}, \bibinfo {author} {\bibfnamefont {F.}~\bibnamefont {Kuemmeth}},
  \bibinfo {author} {\bibfnamefont {G.~A.}\ \bibnamefont {Steele}}, \bibinfo
  {author} {\bibfnamefont {K.}~\bibnamefont {Grove-Rasmussen}}, \bibinfo
  {author} {\bibfnamefont {J.}~\bibnamefont {Nyg\aa{}rd}}, \bibinfo {author}
  {\bibfnamefont {K.}~\bibnamefont {Flensberg}},\ and\ \bibinfo {author}
  {\bibfnamefont {L.~P.}\ \bibnamefont {Kouwenhoven}},\ }\bibfield  {title}
  {\bibinfo {title} {Quantum transport in carbon nanotubes},\ }\href
  {https://doi.org/10.1103/RevModPhys.87.703} {\bibfield  {journal} {\bibinfo
  {journal} {Rev. Mod. Phys.}\ }\textbf {\bibinfo {volume} {87}},\ \bibinfo
  {pages} {703} (\bibinfo {year} {2015})}\BibitemShut {NoStop}%
\bibitem [{\citenamefont {Geim}\ and\ \citenamefont
  {Novoselov}(2007)}]{Geim2007}%
  \BibitemOpen
  \bibfield  {author} {\bibinfo {author} {\bibfnamefont {A.~K.}\ \bibnamefont
  {Geim}}\ and\ \bibinfo {author} {\bibfnamefont {K.~S.}\ \bibnamefont
  {Novoselov}},\ }\bibfield  {title} {\bibinfo {title} {The rise of graphene},\
  }\href {https://doi.org/https://doi.org/10.1038/nmat1849} {\bibfield
  {journal} {\bibinfo  {journal} {Nature Mat.}\ }\textbf {\bibinfo {volume}
  {6}},\ \bibinfo {pages} {183} (\bibinfo {year} {2007})}\BibitemShut {NoStop}%
\bibitem [{\citenamefont {Neto}\ \emph {et~al.}(2009)\citenamefont {Neto},
  \citenamefont {Guinea}, \citenamefont {Peres}, \citenamefont {Novoselov},\
  and\ \citenamefont {Geim}}]{CastroNeto2009}%
  \BibitemOpen
  \bibfield  {author} {\bibinfo {author} {\bibfnamefont {A.~H.~C.}\
  \bibnamefont {Neto}}, \bibinfo {author} {\bibfnamefont {F.}~\bibnamefont
  {Guinea}}, \bibinfo {author} {\bibfnamefont {N.~M.~R.}\ \bibnamefont
  {Peres}}, \bibinfo {author} {\bibfnamefont {K.~S.}\ \bibnamefont
  {Novoselov}},\ and\ \bibinfo {author} {\bibfnamefont {A.~K.}\ \bibnamefont
  {Geim}},\ }\bibfield  {title} {\bibinfo {title} {The electronic properties of
  graphene},\ }\href
  {https://doi.org/https://doi.org/10.1103/RevModPhys.81.109} {\bibfield
  {journal} {\bibinfo  {journal} {Rev. Mod. Phys.}\ }\textbf {\bibinfo {volume}
  {81}},\ \bibinfo {pages} {109} (\bibinfo {year} {2009})}\BibitemShut
  {NoStop}%
\bibitem [{\citenamefont {Abergel}\ \emph {et~al.}(2010)\citenamefont
  {Abergel}, \citenamefont {Apalkov}, \citenamefont {Berashevich},
  \citenamefont {Ziegler},\ and\ \citenamefont {Chakraborty}}]{Abergel2010}%
  \BibitemOpen
  \bibfield  {author} {\bibinfo {author} {\bibfnamefont {D.}~\bibnamefont
  {Abergel}}, \bibinfo {author} {\bibfnamefont {V.}~\bibnamefont {Apalkov}},
  \bibinfo {author} {\bibfnamefont {J.}~\bibnamefont {Berashevich}}, \bibinfo
  {author} {\bibfnamefont {K.}~\bibnamefont {Ziegler}},\ and\ \bibinfo {author}
  {\bibfnamefont {T.}~\bibnamefont {Chakraborty}},\ }\bibfield  {title}
  {\bibinfo {title} {Properties of graphene: a theoretical perspective},\
  }\href {https://doi.org/10.1080/00018732.2010.487978} {\bibfield  {journal}
  {\bibinfo  {journal} {Advances in Physics}\ }\textbf {\bibinfo {volume}
  {59}},\ \bibinfo {pages} {261} (\bibinfo {year} {2010})}\BibitemShut
  {NoStop}%
\bibitem [{\citenamefont {Deshpande}\ and\ \citenamefont
  {Bockrath}(2008)}]{Deshpande2008}%
  \BibitemOpen
  \bibfield  {author} {\bibinfo {author} {\bibfnamefont {V.~V.}\ \bibnamefont
  {Deshpande}}\ and\ \bibinfo {author} {\bibfnamefont {M.}~\bibnamefont
  {Bockrath}},\ }\bibfield  {title} {\bibinfo {title} {The one-dimensional
  {W}igner crystal in carbon nanotubes},\ }\href
  {https://doi.org/https://doi.org/10.1038/nphys895} {\bibfield  {journal}
  {\bibinfo  {journal} {Nature Phys.}\ }\textbf {\bibinfo {volume} {4}},\
  \bibinfo {pages} {314} (\bibinfo {year} {2008})}\BibitemShut {NoStop}%
\bibitem [{\citenamefont {Secchi}\ and\ \citenamefont
  {Rontani}(2009)}]{Secchi2009}%
  \BibitemOpen
  \bibfield  {author} {\bibinfo {author} {\bibfnamefont {A.}~\bibnamefont
  {Secchi}}\ and\ \bibinfo {author} {\bibfnamefont {M.}~\bibnamefont
  {Rontani}},\ }\bibfield  {title} {\bibinfo {title} {Coulomb versus spin-orbit
  interaction in few-electron carbon-nanotube quantum dots},\ }\href
  {https://doi.org/10.1103/PhysRevB.80.041404} {\bibfield  {journal} {\bibinfo
  {journal} {Phys. Rev. B}\ }\textbf {\bibinfo {volume} {80}},\ \bibinfo
  {pages} {041404} (\bibinfo {year} {2009})}\BibitemShut {NoStop}%
\bibitem [{\citenamefont {Secchi}\ and\ \citenamefont
  {Rontani}(2010)}]{Secchi2010}%
  \BibitemOpen
  \bibfield  {author} {\bibinfo {author} {\bibfnamefont {A.}~\bibnamefont
  {Secchi}}\ and\ \bibinfo {author} {\bibfnamefont {M.}~\bibnamefont
  {Rontani}},\ }\bibfield  {title} {\bibinfo {title} {Wigner molecules in
  carbon-nanotube quantum dots},\ }\href
  {https://doi.org/10.1103/PhysRevB.82.035417} {\bibfield  {journal} {\bibinfo
  {journal} {Phys. Rev. B}\ }\textbf {\bibinfo {volume} {82}},\ \bibinfo
  {pages} {035417} (\bibinfo {year} {2010})}\BibitemShut {NoStop}%
\bibitem [{\citenamefont {Pecker}\ \emph {et~al.}(2013)\citenamefont {Pecker},
  \citenamefont {Kuemmeth}, \citenamefont {Secchi}, \citenamefont {Rontani},
  \citenamefont {Ralph}, \citenamefont {McEuen},\ and\ \citenamefont
  {Ilani}}]{Pecker2013}%
  \BibitemOpen
  \bibfield  {author} {\bibinfo {author} {\bibfnamefont {S.}~\bibnamefont
  {Pecker}}, \bibinfo {author} {\bibfnamefont {F.}~\bibnamefont {Kuemmeth}},
  \bibinfo {author} {\bibfnamefont {A.}~\bibnamefont {Secchi}}, \bibinfo
  {author} {\bibfnamefont {M.}~\bibnamefont {Rontani}}, \bibinfo {author}
  {\bibfnamefont {D.~C.}\ \bibnamefont {Ralph}}, \bibinfo {author}
  {\bibfnamefont {P.~L.}\ \bibnamefont {McEuen}},\ and\ \bibinfo {author}
  {\bibfnamefont {S.}~\bibnamefont {Ilani}},\ }\bibfield  {title} {\bibinfo
  {title} {Observation and spectroscopy of a two-electron {W}igner molecule in
  an ultraclean carbon nanotube},\ }\href
  {https://doi.org/https://doi.org/10.1038/nphys2692} {\bibfield  {journal}
  {\bibinfo  {journal} {Nature Phys.}\ }\textbf {\bibinfo {volume} {9}},\
  \bibinfo {pages} {576} (\bibinfo {year} {2013})}\BibitemShut {NoStop}%
\bibitem [{\citenamefont {Shapir}\ \emph {et~al.}(2019)\citenamefont {Shapir},
  \citenamefont {Amo}, \citenamefont {Pecker}, \citenamefont {Moca},
  \citenamefont {Legeza}, \citenamefont {Zarand},\ and\ \citenamefont
  {Ilani}}]{Ilanit2019}%
  \BibitemOpen
  \bibfield  {author} {\bibinfo {author} {\bibfnamefont {I.}~\bibnamefont
  {Shapir}}, \bibinfo {author} {\bibfnamefont {A.}~\bibnamefont {Amo}},
  \bibinfo {author} {\bibfnamefont {S.}~\bibnamefont {Pecker}}, \bibinfo
  {author} {\bibfnamefont {C.~P.}\ \bibnamefont {Moca}}, \bibinfo {author}
  {\bibfnamefont {{\"O}.}~\bibnamefont {Legeza}}, \bibinfo {author}
  {\bibfnamefont {G.}~\bibnamefont {Zarand}},\ and\ \bibinfo {author}
  {\bibfnamefont {S.}~\bibnamefont {Ilani}},\ }\bibfield  {title} {\bibinfo
  {title} {Imaging the electronic {W}igner crystal in one dimension},\ }\href
  {https://doi.org/10.1126/science.aat0905} {\bibfield  {journal} {\bibinfo
  {journal} {Science}\ }\textbf {\bibinfo {volume} {364}},\ \bibinfo {pages}
  {870} (\bibinfo {year} {2019})}\BibitemShut {NoStop}%
\bibitem [{\citenamefont {Lotfizadeh}\ \emph {et~al.}(2019)\citenamefont
  {Lotfizadeh}, \citenamefont {McCulley}, \citenamefont {Senger}, \citenamefont
  {Fu}, \citenamefont {Minot}, \citenamefont {Skinner},\ and\ \citenamefont
  {Deshpande}}]{Lotfizadeh2019}%
  \BibitemOpen
  \bibfield  {author} {\bibinfo {author} {\bibfnamefont {N.}~\bibnamefont
  {Lotfizadeh}}, \bibinfo {author} {\bibfnamefont {D.~R.}\ \bibnamefont
  {McCulley}}, \bibinfo {author} {\bibfnamefont {M.~J.}\ \bibnamefont
  {Senger}}, \bibinfo {author} {\bibfnamefont {H.}~\bibnamefont {Fu}}, \bibinfo
  {author} {\bibfnamefont {E.~D.}\ \bibnamefont {Minot}}, \bibinfo {author}
  {\bibfnamefont {B.}~\bibnamefont {Skinner}},\ and\ \bibinfo {author}
  {\bibfnamefont {V.~V.}\ \bibnamefont {Deshpande}},\ }\bibfield  {title}
  {\bibinfo {title} {Band-gap-dependent electronic compressibility of carbon
  nanotubes in the {Wigner} crystal regime},\ }\href
  {https://doi.org/10.1103/PhysRevLett.123.197701} {\bibfield  {journal}
  {\bibinfo  {journal} {Phys. Rev. Lett.}\ }\textbf {\bibinfo {volume} {123}},\
  \bibinfo {pages} {197701} (\bibinfo {year} {2019})}\BibitemShut {NoStop}%
\bibitem [{\citenamefont {Ziani}\ \emph {et~al.}(2021)\citenamefont {Ziani},
  \citenamefont {Cavaliere}, \citenamefont {Becerra},\ and\ \citenamefont
  {Sassetti}}]{Ziani2021}%
  \BibitemOpen
  \bibfield  {author} {\bibinfo {author} {\bibfnamefont {N.~T.}\ \bibnamefont
  {Ziani}}, \bibinfo {author} {\bibfnamefont {F.}~\bibnamefont {Cavaliere}},
  \bibinfo {author} {\bibfnamefont {K.~G.}\ \bibnamefont {Becerra}},\ and\
  \bibinfo {author} {\bibfnamefont {M.}~\bibnamefont {Sassetti}},\ }\bibfield
  {title} {\bibinfo {title} {A short review of one-dimensional {W}igner
  crystallization},\ }\href {https://doi.org/10.3390/cryst11010020} {\bibfield
  {journal} {\bibinfo  {journal} {Crystals}\ }\textbf {\bibinfo {volume}
  {11}},\ \bibinfo {pages} {20} (\bibinfo {year} {2021})}\BibitemShut {NoStop}%
\bibitem [{\citenamefont {Giamarchi}(2004)}]{Giamarchi2004}%
  \BibitemOpen
  \bibfield  {author} {\bibinfo {author} {\bibfnamefont {T.}~\bibnamefont
  {Giamarchi}},\ }\href@noop {} {\emph {\bibinfo {title} {Quantum Systems in
  One Dimension}}}\ (\bibinfo  {publisher} {Clarendon Press},\ \bibinfo
  {address} {Oxford (UK)},\ \bibinfo {year} {2004})\BibitemShut {NoStop}%
\bibitem [{\citenamefont {Balents}\ and\ \citenamefont
  {Fisher}(1997)}]{Balents1997}%
  \BibitemOpen
  \bibfield  {author} {\bibinfo {author} {\bibfnamefont {L.}~\bibnamefont
  {Balents}}\ and\ \bibinfo {author} {\bibfnamefont {M.~P.~A.}\ \bibnamefont
  {Fisher}},\ }\bibfield  {title} {\bibinfo {title} {Correlation effects in
  carbon nanotubes},\ }\href {https://doi.org/10.1103/PhysRevB.55.R11973}
  {\bibfield  {journal} {\bibinfo  {journal} {Phys. Rev. B}\ }\textbf {\bibinfo
  {volume} {55}},\ \bibinfo {pages} {11973(R)} (\bibinfo {year}
  {1997})}\BibitemShut {NoStop}%
\bibitem [{\citenamefont {Kane}\ \emph {et~al.}(1997)\citenamefont {Kane},
  \citenamefont {Balents},\ and\ \citenamefont {Fisher}}]{Kane1997b}%
  \BibitemOpen
  \bibfield  {author} {\bibinfo {author} {\bibfnamefont {C.~L.}\ \bibnamefont
  {Kane}}, \bibinfo {author} {\bibfnamefont {L.}~\bibnamefont {Balents}},\ and\
  \bibinfo {author} {\bibfnamefont {M.}~\bibnamefont {Fisher}},\ }\bibfield
  {title} {\bibinfo {title} {Coulomb interaction and mesoscopic effects in
  carbon nanotubes},\ }\href {https://doi.org/10.1103/PhysRevLett.79.5086}
  {\bibfield  {journal} {\bibinfo  {journal} {Phys. Rev. Lett.}\ }\textbf
  {\bibinfo {volume} {79}},\ \bibinfo {pages} {5086} (\bibinfo {year}
  {1997})}\BibitemShut {NoStop}%
\bibitem [{\citenamefont {Egger}\ and\ \citenamefont
  {Gogolin}(1997)}]{Egger1997}%
  \BibitemOpen
  \bibfield  {author} {\bibinfo {author} {\bibfnamefont {R.}~\bibnamefont
  {Egger}}\ and\ \bibinfo {author} {\bibfnamefont {A.~O.}\ \bibnamefont
  {Gogolin}},\ }\bibfield  {title} {\bibinfo {title} {Effective low-energy
  theory for correlated carbon nanotubes},\ }\href
  {https://doi.org/10.1103/PhysRevLett.79.5082} {\bibfield  {journal} {\bibinfo
   {journal} {Phys. Rev. Lett.}\ }\textbf {\bibinfo {volume} {79}},\ \bibinfo
  {pages} {5082} (\bibinfo {year} {1997})}\BibitemShut {NoStop}%
\bibitem [{\citenamefont {Krotov}\ \emph {et~al.}(1997)\citenamefont {Krotov},
  \citenamefont {Lee},\ and\ \citenamefont {Louie}}]{Krotov1997}%
  \BibitemOpen
  \bibfield  {author} {\bibinfo {author} {\bibfnamefont {Y.~A.}\ \bibnamefont
  {Krotov}}, \bibinfo {author} {\bibfnamefont {D.}~\bibnamefont {Lee}},\ and\
  \bibinfo {author} {\bibfnamefont {S.~G.}\ \bibnamefont {Louie}},\ }\bibfield
  {title} {\bibinfo {title} {Low energy properties of $(n,n)$ carbon
  nanotubes},\ }\href {https://doi.org/10.1103/PhysRevLett.78.4245} {\bibfield
  {journal} {\bibinfo  {journal} {Phys. Rev. Lett.}\ }\textbf {\bibinfo
  {volume} {78}},\ \bibinfo {pages} {4245} (\bibinfo {year}
  {1997})}\BibitemShut {NoStop}%
\bibitem [{\citenamefont {Yoshioka}\ and\ \citenamefont
  {Odintsov}(1999)}]{Yoshioka1999}%
  \BibitemOpen
  \bibfield  {author} {\bibinfo {author} {\bibfnamefont {H.}~\bibnamefont
  {Yoshioka}}\ and\ \bibinfo {author} {\bibfnamefont {A.~A.}\ \bibnamefont
  {Odintsov}},\ }\bibfield  {title} {\bibinfo {title} {Electronic properties of
  armchair carbon nanotubes: Bosonization approach},\ }\href
  {https://doi.org/10.1103/PhysRevLett.82.374} {\bibfield  {journal} {\bibinfo
  {journal} {Phys. Rev. Lett.}\ }\textbf {\bibinfo {volume} {82}},\ \bibinfo
  {pages} {374} (\bibinfo {year} {1999})}\BibitemShut {NoStop}%
\bibitem [{\citenamefont {Bockrath}\ \emph {et~al.}(1999)\citenamefont
  {Bockrath}, \citenamefont {Cobden}, \citenamefont {Lu}, \citenamefont
  {Rinzler}, \citenamefont {Smalley}, \citenamefont {Balents},\ and\
  \citenamefont {McEuen}}]{Bockrath1999}%
  \BibitemOpen
  \bibfield  {author} {\bibinfo {author} {\bibfnamefont {M.}~\bibnamefont
  {Bockrath}}, \bibinfo {author} {\bibfnamefont {D.~H.}\ \bibnamefont
  {Cobden}}, \bibinfo {author} {\bibfnamefont {J.}~\bibnamefont {Lu}}, \bibinfo
  {author} {\bibfnamefont {A.~G.}\ \bibnamefont {Rinzler}}, \bibinfo {author}
  {\bibfnamefont {R.~E.}\ \bibnamefont {Smalley}}, \bibinfo {author}
  {\bibfnamefont {L.}~\bibnamefont {Balents}},\ and\ \bibinfo {author}
  {\bibfnamefont {P.~L.}\ \bibnamefont {McEuen}},\ }\bibfield  {title}
  {\bibinfo {title} {Luttinger-liquid behaviour in carbon nanotubes},\ }\href
  {https://doi.org/10.1038/17569} {\bibfield  {journal} {\bibinfo  {journal}
  {Nature}\ }\textbf {\bibinfo {volume} {397}},\ \bibinfo {pages} {598}
  (\bibinfo {year} {1999})}\BibitemShut {NoStop}%
\bibitem [{\citenamefont {Postma}\ \emph {et~al.}(2001)\citenamefont {Postma},
  \citenamefont {Teepen}, \citenamefont {Yao}, \citenamefont {Grifoni},\ and\
  \citenamefont {Dekker}}]{Postma2001}%
  \BibitemOpen
  \bibfield  {author} {\bibinfo {author} {\bibfnamefont {H.~W.~C.}\
  \bibnamefont {Postma}}, \bibinfo {author} {\bibfnamefont {T.}~\bibnamefont
  {Teepen}}, \bibinfo {author} {\bibfnamefont {Z.}~\bibnamefont {Yao}},
  \bibinfo {author} {\bibfnamefont {M.}~\bibnamefont {Grifoni}},\ and\ \bibinfo
  {author} {\bibfnamefont {C.}~\bibnamefont {Dekker}},\ }\bibfield  {title}
  {\bibinfo {title} {Carbon nanotube single-electron transistors at room
  temperature},\ }\href {https://doi.org/10.1126/science.1061797} {\bibfield
  {journal} {\bibinfo  {journal} {Science}\ }\textbf {\bibinfo {volume}
  {293}},\ \bibinfo {pages} {76} (\bibinfo {year} {2001})}\BibitemShut
  {NoStop}%
\bibitem [{\citenamefont {Cao}\ \emph {et~al.}(2018{\natexlab{a}})\citenamefont
  {Cao}, \citenamefont {Fatemi}, \citenamefont {Fang}, \citenamefont
  {Watanabe}, \citenamefont {Taniguchi}, \citenamefont {Kaxiras},\ and\
  \citenamefont {Jarillo-Herrero}}]{Cao2018}%
  \BibitemOpen
  \bibfield  {author} {\bibinfo {author} {\bibfnamefont {Y.}~\bibnamefont
  {Cao}}, \bibinfo {author} {\bibfnamefont {V.}~\bibnamefont {Fatemi}},
  \bibinfo {author} {\bibfnamefont {S.}~\bibnamefont {Fang}}, \bibinfo {author}
  {\bibfnamefont {K.}~\bibnamefont {Watanabe}}, \bibinfo {author}
  {\bibfnamefont {T.}~\bibnamefont {Taniguchi}}, \bibinfo {author}
  {\bibfnamefont {E.}~\bibnamefont {Kaxiras}},\ and\ \bibinfo {author}
  {\bibfnamefont {P.}~\bibnamefont {Jarillo-Herrero}},\ }\bibfield  {title}
  {\bibinfo {title} {Unconventional superconductivity in magic-angle graphene
  superlattices},\ }\href {https://doi.org/10.1038/nature26160} {\bibfield
  {journal} {\bibinfo  {journal} {Nature}\ }\textbf {\bibinfo {volume} {556}},\
  \bibinfo {pages} {43} (\bibinfo {year} {2018}{\natexlab{a}})}\BibitemShut
  {NoStop}%
\bibitem [{\citenamefont {Yankowitz}\ \emph {et~al.}(2019)\citenamefont
  {Yankowitz}, \citenamefont {Chen}, \citenamefont {Polshyn}, \citenamefont
  {Zhang}, \citenamefont {Watanabe}, \citenamefont {Taniguchi}, \citenamefont
  {Graf}, \citenamefont {Young},\ and\ \citenamefont {Dean}}]{Yankowitz2019}%
  \BibitemOpen
  \bibfield  {author} {\bibinfo {author} {\bibfnamefont {M.}~\bibnamefont
  {Yankowitz}}, \bibinfo {author} {\bibfnamefont {S.}~\bibnamefont {Chen}},
  \bibinfo {author} {\bibfnamefont {H.}~\bibnamefont {Polshyn}}, \bibinfo
  {author} {\bibfnamefont {Y.}~\bibnamefont {Zhang}}, \bibinfo {author}
  {\bibfnamefont {K.}~\bibnamefont {Watanabe}}, \bibinfo {author}
  {\bibfnamefont {T.}~\bibnamefont {Taniguchi}}, \bibinfo {author}
  {\bibfnamefont {D.}~\bibnamefont {Graf}}, \bibinfo {author} {\bibfnamefont
  {A.~F.}\ \bibnamefont {Young}},\ and\ \bibinfo {author} {\bibfnamefont
  {C.~R.}\ \bibnamefont {Dean}},\ }\bibfield  {title} {\bibinfo {title} {Tuning
  superconductivity in twisted bilayer graphene},\ }\href
  {https://doi.org/10.1126/science.aav1910} {\bibfield  {journal} {\bibinfo
  {journal} {Science}\ }\textbf {\bibinfo {volume} {363}},\ \bibinfo {pages}
  {1059} (\bibinfo {year} {2019})}\BibitemShut {NoStop}%
\bibitem [{\citenamefont {Hao}\ \emph {et~al.}(2021)\citenamefont {Hao},
  \citenamefont {Zimmerman}, \citenamefont {Ledwith}, \citenamefont {Khalaf},
  \citenamefont {Najafabadi}, \citenamefont {Watanabe}, \citenamefont
  {Taniguchi}, \citenamefont {Vishwanath},\ and\ \citenamefont
  {Kim}}]{Hao2021}%
  \BibitemOpen
  \bibfield  {author} {\bibinfo {author} {\bibfnamefont {Z.}~\bibnamefont
  {Hao}}, \bibinfo {author} {\bibfnamefont {A.~M.}\ \bibnamefont {Zimmerman}},
  \bibinfo {author} {\bibfnamefont {P.}~\bibnamefont {Ledwith}}, \bibinfo
  {author} {\bibfnamefont {E.}~\bibnamefont {Khalaf}}, \bibinfo {author}
  {\bibfnamefont {D.~H.}\ \bibnamefont {Najafabadi}}, \bibinfo {author}
  {\bibfnamefont {K.}~\bibnamefont {Watanabe}}, \bibinfo {author}
  {\bibfnamefont {T.}~\bibnamefont {Taniguchi}}, \bibinfo {author}
  {\bibfnamefont {A.}~\bibnamefont {Vishwanath}},\ and\ \bibinfo {author}
  {\bibfnamefont {P.}~\bibnamefont {Kim}},\ }\bibfield  {title} {\bibinfo
  {title} {Electric field-tunable superconductivity in alternating-twist
  magic-angle trilayer graphene},\ }\href
  {https://doi.org/10.1126/science.abg0399} {\bibfield  {journal} {\bibinfo
  {journal} {Science}\ }\textbf {\bibinfo {volume} {371}},\ \bibinfo {pages}
  {1133} (\bibinfo {year} {2021})}\BibitemShut {NoStop}%
\bibitem [{\citenamefont {Park}\ \emph {et~al.}(2021)\citenamefont {Park},
  \citenamefont {Cao}, \citenamefont {Watanabe}, \citenamefont {Taniguchi},\
  and\ \citenamefont {Jarillo-Herrero}}]{Park2021}%
  \BibitemOpen
  \bibfield  {author} {\bibinfo {author} {\bibfnamefont {J.~M.}\ \bibnamefont
  {Park}}, \bibinfo {author} {\bibfnamefont {Y.}~\bibnamefont {Cao}}, \bibinfo
  {author} {\bibfnamefont {K.}~\bibnamefont {Watanabe}}, \bibinfo {author}
  {\bibfnamefont {T.}~\bibnamefont {Taniguchi}},\ and\ \bibinfo {author}
  {\bibfnamefont {P.}~\bibnamefont {Jarillo-Herrero}},\ }\bibfield  {title}
  {\bibinfo {title} {Tunable strongly coupled superconductivity in magic-angle
  twisted trilayer graphene},\ }\bibfield  {journal} {\bibinfo  {journal}
  {Nature}\ }\textbf {\bibinfo {volume} {590}},\ \href
  {https://doi.org/10.1038/s41586-021-03192-0} {10.1038/s41586-021-03192-0}
  (\bibinfo {year} {2021})\BibitemShut {NoStop}%
\bibitem [{\citenamefont {Zhou}\ \emph {et~al.}(2022)\citenamefont {Zhou},
  \citenamefont {Holleis}, \citenamefont {Saito}, \citenamefont {Cohen},
  \citenamefont {Huynh}, \citenamefont {Patterson}, \citenamefont {Yang},
  \citenamefont {Taniguchi}, \citenamefont {Watanabe},\ and\ \citenamefont
  {Young}}]{Zhou2022}%
  \BibitemOpen
  \bibfield  {author} {\bibinfo {author} {\bibfnamefont {H.}~\bibnamefont
  {Zhou}}, \bibinfo {author} {\bibfnamefont {L.}~\bibnamefont {Holleis}},
  \bibinfo {author} {\bibfnamefont {Y.}~\bibnamefont {Saito}}, \bibinfo
  {author} {\bibfnamefont {L.}~\bibnamefont {Cohen}}, \bibinfo {author}
  {\bibfnamefont {W.}~\bibnamefont {Huynh}}, \bibinfo {author} {\bibfnamefont
  {C.~L.}\ \bibnamefont {Patterson}}, \bibinfo {author} {\bibfnamefont
  {F.}~\bibnamefont {Yang}}, \bibinfo {author} {\bibfnamefont {T.}~\bibnamefont
  {Taniguchi}}, \bibinfo {author} {\bibfnamefont {K.}~\bibnamefont
  {Watanabe}},\ and\ \bibinfo {author} {\bibfnamefont {A.~F.}\ \bibnamefont
  {Young}},\ }\bibfield  {title} {\bibinfo {title} {Isospin magnetism and
  spin-polarized superconductivity in bernal bilayer graphene},\ }\href
  {https://doi.org/10.1126/science.abm8386} {\bibfield  {journal} {\bibinfo
  {journal} {Science}\ }\textbf {\bibinfo {volume} {375}},\ \bibinfo {pages}
  {774} (\bibinfo {year} {2022})}\BibitemShut {NoStop}%
\bibitem [{\citenamefont {Liu}\ \emph {et~al.}(2017)\citenamefont {Liu},
  \citenamefont {Watanabe}, \citenamefont {Taniguchi}, \citenamefont
  {Halperin},\ and\ \citenamefont {Kim}}]{Liu2017}%
  \BibitemOpen
  \bibfield  {author} {\bibinfo {author} {\bibfnamefont {X.}~\bibnamefont
  {Liu}}, \bibinfo {author} {\bibfnamefont {K.}~\bibnamefont {Watanabe}},
  \bibinfo {author} {\bibfnamefont {T.}~\bibnamefont {Taniguchi}}, \bibinfo
  {author} {\bibfnamefont {B.~I.}\ \bibnamefont {Halperin}},\ and\ \bibinfo
  {author} {\bibfnamefont {P.}~\bibnamefont {Kim}},\ }\bibfield  {title}
  {\bibinfo {title} {Quantum {H}all drag of exciton condensate in graphene},\
  }\href {https://doi.org/10.1038/NPHYS4116} {\bibfield  {journal} {\bibinfo
  {journal} {Nature Phys.}\ }\textbf {\bibinfo {volume} {13}},\ \bibinfo
  {pages} {746} (\bibinfo {year} {2017})}\BibitemShut {NoStop}%
\bibitem [{\citenamefont {Li}\ \emph {et~al.}(2017)\citenamefont {Li},
  \citenamefont {Taniguchi}, \citenamefont {Watanabe}, \citenamefont {Hone},\
  and\ \citenamefont {Dean}}]{Li2017}%
  \BibitemOpen
  \bibfield  {author} {\bibinfo {author} {\bibfnamefont {J.~I.~A.}\
  \bibnamefont {Li}}, \bibinfo {author} {\bibfnamefont {T.}~\bibnamefont
  {Taniguchi}}, \bibinfo {author} {\bibfnamefont {K.}~\bibnamefont {Watanabe}},
  \bibinfo {author} {\bibfnamefont {J.}~\bibnamefont {Hone}},\ and\ \bibinfo
  {author} {\bibfnamefont {C.~R.}\ \bibnamefont {Dean}},\ }\bibfield  {title}
  {\bibinfo {title} {Excitonic superfluid phase in double bilayer graphene},\
  }\href {https://doi.org/10.1038/NPHYS4140} {\bibfield  {journal} {\bibinfo
  {journal} {Nature Phys.}\ }\textbf {\bibinfo {volume} {13}},\ \bibinfo
  {pages} {751} (\bibinfo {year} {2017})}\BibitemShut {NoStop}%
\bibitem [{\citenamefont {Liu}\ \emph {et~al.}(2022)\citenamefont {Liu},
  \citenamefont {Li}, \citenamefont {Watanabe}, \citenamefont {Taniguchi},
  \citenamefont {Hone}, \citenamefont {Halperin}, \citenamefont {Kim},\ and\
  \citenamefont {Dean}}]{Liu2022}%
  \BibitemOpen
  \bibfield  {author} {\bibinfo {author} {\bibfnamefont {X.}~\bibnamefont
  {Liu}}, \bibinfo {author} {\bibfnamefont {J.~I.~A.}\ \bibnamefont {Li}},
  \bibinfo {author} {\bibfnamefont {K.}~\bibnamefont {Watanabe}}, \bibinfo
  {author} {\bibfnamefont {T.}~\bibnamefont {Taniguchi}}, \bibinfo {author}
  {\bibfnamefont {J.}~\bibnamefont {Hone}}, \bibinfo {author} {\bibfnamefont
  {B.~I.}\ \bibnamefont {Halperin}}, \bibinfo {author} {\bibfnamefont
  {P.}~\bibnamefont {Kim}},\ and\ \bibinfo {author} {\bibfnamefont {C.~R.}\
  \bibnamefont {Dean}},\ }\bibfield  {title} {\bibinfo {title} {Crossover
  between strongly coupled and weakly coupled exciton superfluids},\ }\href
  {https://doi.org/10.1126/science.abg1110} {\bibfield  {journal} {\bibinfo
  {journal} {Science}\ }\textbf {\bibinfo {volume} {375}},\ \bibinfo {pages}
  {205} (\bibinfo {year} {2022})}\BibitemShut {NoStop}%
\bibitem [{\citenamefont {Li}\ \emph {et~al.}(2024)\citenamefont {Li},
  \citenamefont {Chen}, \citenamefont {Wei}, \citenamefont {Chen},
  \citenamefont {Huang}, \citenamefont {Zhu}, \citenamefont {Zhu},
  \citenamefont {An}, \citenamefont {Song}, \citenamefont {Gan}, \citenamefont
  {Zhang}, \citenamefont {Watanabe}, \citenamefont {Taniguchi}, \citenamefont
  {Shi}, \citenamefont {Novoselov}, \citenamefont {Wang}, \citenamefont {Yu},\
  and\ \citenamefont {Wang}}]{Li2024}%
  \BibitemOpen
  \bibfield  {author} {\bibinfo {author} {\bibfnamefont {Q.}~\bibnamefont
  {Li}}, \bibinfo {author} {\bibfnamefont {Y.}~\bibnamefont {Chen}}, \bibinfo
  {author} {\bibfnamefont {L.}~\bibnamefont {Wei}}, \bibinfo {author}
  {\bibfnamefont {H.}~\bibnamefont {Chen}}, \bibinfo {author} {\bibfnamefont
  {Y.}~\bibnamefont {Huang}}, \bibinfo {author} {\bibfnamefont
  {Y.}~\bibnamefont {Zhu}}, \bibinfo {author} {\bibfnamefont {W.}~\bibnamefont
  {Zhu}}, \bibinfo {author} {\bibfnamefont {D.}~\bibnamefont {An}}, \bibinfo
  {author} {\bibfnamefont {J.}~\bibnamefont {Song}}, \bibinfo {author}
  {\bibfnamefont {Q.}~\bibnamefont {Gan}}, \bibinfo {author} {\bibfnamefont
  {Q.}~\bibnamefont {Zhang}}, \bibinfo {author} {\bibfnamefont
  {K.}~\bibnamefont {Watanabe}}, \bibinfo {author} {\bibfnamefont
  {T.}~\bibnamefont {Taniguchi}}, \bibinfo {author} {\bibfnamefont
  {X.}~\bibnamefont {Shi}}, \bibinfo {author} {\bibfnamefont {K.~S.}\
  \bibnamefont {Novoselov}}, \bibinfo {author} {\bibfnamefont {R.}~\bibnamefont
  {Wang}}, \bibinfo {author} {\bibfnamefont {G.}~\bibnamefont {Yu}},\ and\
  \bibinfo {author} {\bibfnamefont {L.}~\bibnamefont {Wang}},\ }\bibfield
  {title} {\bibinfo {title} {Strongly coupled magneto-exciton condensates in
  large-angle twisted double bilayer graphene},\ }\bibfield  {journal}
  {\bibinfo  {journal} {Nature Communications}\ }\textbf {\bibinfo {volume}
  {15}},\ \href {https://doi.org/10.1038/s41467-024-49406-7}
  {10.1038/s41467-024-49406-7} (\bibinfo {year} {2024})\BibitemShut {NoStop}%
\bibitem [{\citenamefont {Cao}\ \emph {et~al.}(2018{\natexlab{b}})\citenamefont
  {Cao}, \citenamefont {Fatemi}, \citenamefont {Demir}, \citenamefont {Fang},
  \citenamefont {Tomarken}, \citenamefont {Luo}, \citenamefont
  {Sanchez-Yamagishi}, \citenamefont {Watanabe}, \citenamefont {Taniguchi},
  \citenamefont {Kaxiras}, \citenamefont {Ashoori},\ and\ \citenamefont
  {Jarillo-Herrero}}]{Cao2018bis}%
  \BibitemOpen
  \bibfield  {author} {\bibinfo {author} {\bibfnamefont {Y.}~\bibnamefont
  {Cao}}, \bibinfo {author} {\bibfnamefont {V.}~\bibnamefont {Fatemi}},
  \bibinfo {author} {\bibfnamefont {A.}~\bibnamefont {Demir}}, \bibinfo
  {author} {\bibfnamefont {S.}~\bibnamefont {Fang}}, \bibinfo {author}
  {\bibfnamefont {S.~L.}\ \bibnamefont {Tomarken}}, \bibinfo {author}
  {\bibfnamefont {J.~Y.}\ \bibnamefont {Luo}}, \bibinfo {author} {\bibfnamefont
  {J.~D.}\ \bibnamefont {Sanchez-Yamagishi}}, \bibinfo {author} {\bibfnamefont
  {K.}~\bibnamefont {Watanabe}}, \bibinfo {author} {\bibfnamefont
  {T.}~\bibnamefont {Taniguchi}}, \bibinfo {author} {\bibfnamefont
  {E.}~\bibnamefont {Kaxiras}}, \bibinfo {author} {\bibfnamefont {R.~C.}\
  \bibnamefont {Ashoori}},\ and\ \bibinfo {author} {\bibfnamefont
  {P.}~\bibnamefont {Jarillo-Herrero}},\ }\bibfield  {title} {\bibinfo {title}
  {Correlated insulator behaviour at half-filling in magic-angle graphene
  superlattices},\ }\href {https://doi.org/10.1038/nature26154} {\bibfield
  {journal} {\bibinfo  {journal} {Nature}\ }\textbf {\bibinfo {volume} {556}},\
  \bibinfo {pages} {80} (\bibinfo {year} {2018}{\natexlab{b}})}\BibitemShut
  {NoStop}%
\bibitem [{\citenamefont {Sharpe}\ \emph {et~al.}(2019)\citenamefont {Sharpe},
  \citenamefont {Fox}, \citenamefont {Barnard}, \citenamefont {Finney},
  \citenamefont {Watanabe}, \citenamefont {Taniguchi}, \citenamefont
  {Kastner},\ and\ \citenamefont {Goldhaber-Gordon}}]{Sharpe2019}%
  \BibitemOpen
  \bibfield  {author} {\bibinfo {author} {\bibfnamefont {A.~L.}\ \bibnamefont
  {Sharpe}}, \bibinfo {author} {\bibfnamefont {E.~J.}\ \bibnamefont {Fox}},
  \bibinfo {author} {\bibfnamefont {A.~W.}\ \bibnamefont {Barnard}}, \bibinfo
  {author} {\bibfnamefont {J.}~\bibnamefont {Finney}}, \bibinfo {author}
  {\bibfnamefont {K.}~\bibnamefont {Watanabe}}, \bibinfo {author}
  {\bibfnamefont {T.}~\bibnamefont {Taniguchi}}, \bibinfo {author}
  {\bibfnamefont {M.~A.}\ \bibnamefont {Kastner}},\ and\ \bibinfo {author}
  {\bibfnamefont {D.}~\bibnamefont {Goldhaber-Gordon}},\ }\bibfield  {title}
  {\bibinfo {title} {Emergent ferromagnetism near three-quarters filling in
  twisted bilayer graphene},\ }\href {https://doi.org/10.1126/science.aaw3780}
  {\bibfield  {journal} {\bibinfo  {journal} {Science}\ }\textbf {\bibinfo
  {volume} {365}},\ \bibinfo {pages} {605} (\bibinfo {year}
  {2019})}\BibitemShut {NoStop}%
\bibitem [{\citenamefont {Stepanov}\ \emph {et~al.}(2020)\citenamefont
  {Stepanov}, \citenamefont {Das}, \citenamefont {Lu}, \citenamefont
  {Fahimniya}, \citenamefont {Watanabe}, \citenamefont {Taniguchi},
  \citenamefont {Koppens}, \citenamefont {Lischner}, \citenamefont {Levitov},\
  and\ \citenamefont {Efetov}}]{Stepanov2020}%
  \BibitemOpen
  \bibfield  {author} {\bibinfo {author} {\bibfnamefont {P.}~\bibnamefont
  {Stepanov}}, \bibinfo {author} {\bibfnamefont {I.}~\bibnamefont {Das}},
  \bibinfo {author} {\bibfnamefont {X.}~\bibnamefont {Lu}}, \bibinfo {author}
  {\bibfnamefont {A.}~\bibnamefont {Fahimniya}}, \bibinfo {author}
  {\bibfnamefont {K.}~\bibnamefont {Watanabe}}, \bibinfo {author}
  {\bibfnamefont {T.}~\bibnamefont {Taniguchi}}, \bibinfo {author}
  {\bibfnamefont {F.~H.~L.}\ \bibnamefont {Koppens}}, \bibinfo {author}
  {\bibfnamefont {J.}~\bibnamefont {Lischner}}, \bibinfo {author}
  {\bibfnamefont {L.}~\bibnamefont {Levitov}},\ and\ \bibinfo {author}
  {\bibfnamefont {D.~K.}\ \bibnamefont {Efetov}},\ }\bibfield  {title}
  {\bibinfo {title} {Untying the insulating and superconducting orders in
  magic-angle graphene},\ }\href {https://doi.org/10.1038/s41586-020-2459-6}
  {\bibfield  {journal} {\bibinfo  {journal} {Nature}\ }\textbf {\bibinfo
  {volume} {583}},\ \bibinfo {pages} {375} (\bibinfo {year}
  {2020})}\BibitemShut {NoStop}%
\bibitem [{\citenamefont {Saito}\ \emph {et~al.}(2020)\citenamefont {Saito},
  \citenamefont {Ge}, \citenamefont {Watanabe}, \citenamefont {Taniguchi},\
  and\ \citenamefont {Young}}]{Saito2020}%
  \BibitemOpen
  \bibfield  {author} {\bibinfo {author} {\bibfnamefont {Y.}~\bibnamefont
  {Saito}}, \bibinfo {author} {\bibfnamefont {J.}~\bibnamefont {Ge}}, \bibinfo
  {author} {\bibfnamefont {K.}~\bibnamefont {Watanabe}}, \bibinfo {author}
  {\bibfnamefont {T.}~\bibnamefont {Taniguchi}},\ and\ \bibinfo {author}
  {\bibfnamefont {A.~F.}\ \bibnamefont {Young}},\ }\bibfield  {title} {\bibinfo
  {title} {Independent superconductors and correlated insulators in twisted
  bilayer graphene},\ }\href {https://doi.org/10.1038/s41567-020-0928-3}
  {\bibfield  {journal} {\bibinfo  {journal} {Nature Physics}\ }\textbf
  {\bibinfo {volume} {16}},\ \bibinfo {pages} {926} (\bibinfo {year}
  {2020})}\BibitemShut {NoStop}%
\bibitem [{\citenamefont {Liu}\ \emph {et~al.}(2020)\citenamefont {Liu},
  \citenamefont {Hao}, \citenamefont {Khalaf}, \citenamefont {Lee},
  \citenamefont {Ronen}, \citenamefont {Yoo}, \citenamefont {Najafabadi},
  \citenamefont {Watanabe}, \citenamefont {Taniguchi}, \citenamefont
  {Vishwanath},\ and\ \citenamefont {Kim}}]{Liu2020}%
  \BibitemOpen
  \bibfield  {author} {\bibinfo {author} {\bibfnamefont {X.}~\bibnamefont
  {Liu}}, \bibinfo {author} {\bibfnamefont {Z.}~\bibnamefont {Hao}}, \bibinfo
  {author} {\bibfnamefont {E.}~\bibnamefont {Khalaf}}, \bibinfo {author}
  {\bibfnamefont {J.~Y.}\ \bibnamefont {Lee}}, \bibinfo {author} {\bibfnamefont
  {Y.}~\bibnamefont {Ronen}}, \bibinfo {author} {\bibfnamefont
  {H.}~\bibnamefont {Yoo}}, \bibinfo {author} {\bibfnamefont {D.~H.}\
  \bibnamefont {Najafabadi}}, \bibinfo {author} {\bibfnamefont
  {K.}~\bibnamefont {Watanabe}}, \bibinfo {author} {\bibfnamefont
  {T.}~\bibnamefont {Taniguchi}}, \bibinfo {author} {\bibfnamefont
  {A.}~\bibnamefont {Vishwanath}},\ and\ \bibinfo {author} {\bibfnamefont
  {P.}~\bibnamefont {Kim}},\ }\bibfield  {title} {\bibinfo {title} {Tunable
  spin-polarized correlated states in twisted double bilayer graphene},\ }\href
  {https://doi.org/10.1038/s41586-020-2458-7} {\bibfield  {journal} {\bibinfo
  {journal} {Nature}\ }\textbf {\bibinfo {volume} {583}},\ \bibinfo {pages}
  {221} (\bibinfo {year} {2020})}\BibitemShut {NoStop}%
\bibitem [{\citenamefont {Shen}\ \emph {et~al.}(2020)\citenamefont {Shen},
  \citenamefont {Chu}, \citenamefont {Wu}, \citenamefont {Li}, \citenamefont
  {Wang}, \citenamefont {Zhao}, \citenamefont {Tang}, \citenamefont {Liu},
  \citenamefont {Tian}, \citenamefont {Watanabe}, \citenamefont {Taniguchi},
  \citenamefont {Yang}, \citenamefont {Meng}, \citenamefont {Shi},
  \citenamefont {Yazyev},\ and\ \citenamefont {Zhang}}]{Shen2020}%
  \BibitemOpen
  \bibfield  {author} {\bibinfo {author} {\bibfnamefont {C.}~\bibnamefont
  {Shen}}, \bibinfo {author} {\bibfnamefont {Y.}~\bibnamefont {Chu}}, \bibinfo
  {author} {\bibfnamefont {Q.}~\bibnamefont {Wu}}, \bibinfo {author}
  {\bibfnamefont {N.}~\bibnamefont {Li}}, \bibinfo {author} {\bibfnamefont
  {S.}~\bibnamefont {Wang}}, \bibinfo {author} {\bibfnamefont {Y.}~\bibnamefont
  {Zhao}}, \bibinfo {author} {\bibfnamefont {J.}~\bibnamefont {Tang}}, \bibinfo
  {author} {\bibfnamefont {J.}~\bibnamefont {Liu}}, \bibinfo {author}
  {\bibfnamefont {J.}~\bibnamefont {Tian}}, \bibinfo {author} {\bibfnamefont
  {K.}~\bibnamefont {Watanabe}}, \bibinfo {author} {\bibfnamefont
  {T.}~\bibnamefont {Taniguchi}}, \bibinfo {author} {\bibfnamefont
  {R.}~\bibnamefont {Yang}}, \bibinfo {author} {\bibfnamefont {Z.~Y.}\
  \bibnamefont {Meng}}, \bibinfo {author} {\bibfnamefont {D.}~\bibnamefont
  {Shi}}, \bibinfo {author} {\bibfnamefont {O.}~\bibnamefont {Yazyev},
  \bibfnamefont {V}},\ and\ \bibinfo {author} {\bibfnamefont {G.}~\bibnamefont
  {Zhang}},\ }\bibfield  {title} {\bibinfo {title} {Correlated states in
  twisted double bilayer graphene},\ }\href
  {https://doi.org/10.1038/s41567-020-0825-9} {\bibfield  {journal} {\bibinfo
  {journal} {Nature Physics}\ }\textbf {\bibinfo {volume} {16}},\ \bibinfo
  {pages} {520} (\bibinfo {year} {2020})}\BibitemShut {NoStop}%
\bibitem [{\citenamefont {He}\ \emph {et~al.}(2021)\citenamefont {He},
  \citenamefont {Li}, \citenamefont {Cai}, \citenamefont {Liu}, \citenamefont
  {Watanabe}, \citenamefont {Taniguchi}, \citenamefont {Xu},\ and\
  \citenamefont {Yankowitz}}]{He2021}%
  \BibitemOpen
  \bibfield  {author} {\bibinfo {author} {\bibfnamefont {M.}~\bibnamefont
  {He}}, \bibinfo {author} {\bibfnamefont {Y.}~\bibnamefont {Li}}, \bibinfo
  {author} {\bibfnamefont {J.}~\bibnamefont {Cai}}, \bibinfo {author}
  {\bibfnamefont {Y.}~\bibnamefont {Liu}}, \bibinfo {author} {\bibfnamefont
  {K.}~\bibnamefont {Watanabe}}, \bibinfo {author} {\bibfnamefont
  {T.}~\bibnamefont {Taniguchi}}, \bibinfo {author} {\bibfnamefont
  {X.}~\bibnamefont {Xu}},\ and\ \bibinfo {author} {\bibfnamefont
  {M.}~\bibnamefont {Yankowitz}},\ }\bibfield  {title} {\bibinfo {title}
  {Symmetry breaking in twisted double bilayer graphene},\ }\href
  {https://doi.org/10.1038/s41567-020-1030-6} {\bibfield  {journal} {\bibinfo
  {journal} {Nature Physics}\ }\textbf {\bibinfo {volume} {17}},\ \bibinfo
  {pages} {26} (\bibinfo {year} {2021})}\BibitemShut {NoStop}%
\bibitem [{\citenamefont {Chen}\ \emph {et~al.}(2021)\citenamefont {Chen},
  \citenamefont {He}, \citenamefont {Zhang}, \citenamefont {Hsieh},
  \citenamefont {Fei}, \citenamefont {Watanabe}, \citenamefont {Taniguchi},
  \citenamefont {Cobden}, \citenamefont {Xu}, \citenamefont {Dean},\ and\
  \citenamefont {Yankowitz}}]{Chen2021}%
  \BibitemOpen
  \bibfield  {author} {\bibinfo {author} {\bibfnamefont {S.}~\bibnamefont
  {Chen}}, \bibinfo {author} {\bibfnamefont {M.}~\bibnamefont {He}}, \bibinfo
  {author} {\bibfnamefont {Y.-H.}\ \bibnamefont {Zhang}}, \bibinfo {author}
  {\bibfnamefont {V.}~\bibnamefont {Hsieh}}, \bibinfo {author} {\bibfnamefont
  {Z.}~\bibnamefont {Fei}}, \bibinfo {author} {\bibfnamefont {K.}~\bibnamefont
  {Watanabe}}, \bibinfo {author} {\bibfnamefont {T.}~\bibnamefont {Taniguchi}},
  \bibinfo {author} {\bibfnamefont {D.~H.}\ \bibnamefont {Cobden}}, \bibinfo
  {author} {\bibfnamefont {X.}~\bibnamefont {Xu}}, \bibinfo {author}
  {\bibfnamefont {C.~R.}\ \bibnamefont {Dean}},\ and\ \bibinfo {author}
  {\bibfnamefont {M.}~\bibnamefont {Yankowitz}},\ }\bibfield  {title} {\bibinfo
  {title} {Electrically tunable correlated and topological states in twisted
  monolayer-bilayer graphene},\ }\href
  {https://doi.org/10.1038/s41567-020-01062-6} {\bibfield  {journal} {\bibinfo
  {journal} {Nature Physics}\ }\textbf {\bibinfo {volume} {17}},\ \bibinfo
  {pages} {374} (\bibinfo {year} {2021})}\BibitemShut {NoStop}%
\bibitem [{\citenamefont {Rickhaus}\ \emph {et~al.}(2021)\citenamefont
  {Rickhaus}, \citenamefont {de~Vries}, \citenamefont {Zhu}, \citenamefont
  {Portoles}, \citenamefont {Zheng}, \citenamefont {Masseroni}, \citenamefont
  {Kurzmann}, \citenamefont {Taniguchi}, \citenamefont {Watanabe},
  \citenamefont {MacDonald}, \citenamefont {Ihn},\ and\ \citenamefont
  {Ensslin}}]{Rickhaus2021}%
  \BibitemOpen
  \bibfield  {author} {\bibinfo {author} {\bibfnamefont {P.}~\bibnamefont
  {Rickhaus}}, \bibinfo {author} {\bibfnamefont {F.~K.}\ \bibnamefont
  {de~Vries}}, \bibinfo {author} {\bibfnamefont {J.}~\bibnamefont {Zhu}},
  \bibinfo {author} {\bibfnamefont {E.}~\bibnamefont {Portoles}}, \bibinfo
  {author} {\bibfnamefont {G.}~\bibnamefont {Zheng}}, \bibinfo {author}
  {\bibfnamefont {M.}~\bibnamefont {Masseroni}}, \bibinfo {author}
  {\bibfnamefont {A.}~\bibnamefont {Kurzmann}}, \bibinfo {author}
  {\bibfnamefont {T.}~\bibnamefont {Taniguchi}}, \bibinfo {author}
  {\bibfnamefont {K.}~\bibnamefont {Watanabe}}, \bibinfo {author}
  {\bibfnamefont {A.~H.}\ \bibnamefont {MacDonald}}, \bibinfo {author}
  {\bibfnamefont {T.}~\bibnamefont {Ihn}},\ and\ \bibinfo {author}
  {\bibfnamefont {K.}~\bibnamefont {Ensslin}},\ }\bibfield  {title} {\bibinfo
  {title} {Correlated electron-hole state in twisted double-bilayer graphene},\
  }\href {https://doi.org/10.1126/science.abc3534} {\bibfield  {journal}
  {\bibinfo  {journal} {Science}\ }\textbf {\bibinfo {volume} {373}},\ \bibinfo
  {pages} {1257} (\bibinfo {year} {2021})}\BibitemShut {NoStop}%
\bibitem [{\citenamefont {Deshpande}\ \emph {et~al.}(2009)\citenamefont
  {Deshpande}, \citenamefont {Chandra}, \citenamefont {Caldwell}, \citenamefont
  {Novikov}, \citenamefont {Hone},\ and\ \citenamefont
  {Bockrath}}]{Deshpande2009}%
  \BibitemOpen
  \bibfield  {author} {\bibinfo {author} {\bibfnamefont {V.~V.}\ \bibnamefont
  {Deshpande}}, \bibinfo {author} {\bibfnamefont {B.}~\bibnamefont {Chandra}},
  \bibinfo {author} {\bibfnamefont {R.}~\bibnamefont {Caldwell}}, \bibinfo
  {author} {\bibfnamefont {D.~S.}\ \bibnamefont {Novikov}}, \bibinfo {author}
  {\bibfnamefont {J.}~\bibnamefont {Hone}},\ and\ \bibinfo {author}
  {\bibfnamefont {M.}~\bibnamefont {Bockrath}},\ }\bibfield  {title} {\bibinfo
  {title} {Mott insulating state in ultraclean carbon nanotubes},\ }\href
  {https://doi.org/10.1126/science.1165799} {\bibfield  {journal} {\bibinfo
  {journal} {Science}\ }\textbf {\bibinfo {volume} {323}},\ \bibinfo {pages}
  {106} (\bibinfo {year} {2009})}\BibitemShut {NoStop}%
\bibitem [{\citenamefont {Senger}\ \emph {et~al.}(2018)\citenamefont {Senger},
  \citenamefont {McCulley}, \citenamefont {Lotfizadeh}, \citenamefont
  {Deshpande},\ and\ \citenamefont {Minot}}]{Senger2018}%
  \BibitemOpen
  \bibfield  {author} {\bibinfo {author} {\bibfnamefont {M.~J.}\ \bibnamefont
  {Senger}}, \bibinfo {author} {\bibfnamefont {D.~R.}\ \bibnamefont
  {McCulley}}, \bibinfo {author} {\bibfnamefont {N.}~\bibnamefont
  {Lotfizadeh}}, \bibinfo {author} {\bibfnamefont {V.~V.}\ \bibnamefont
  {Deshpande}},\ and\ \bibinfo {author} {\bibfnamefont {E.~D.}\ \bibnamefont
  {Minot}},\ }\bibfield  {title} {\bibinfo {title} {Universal
  interaction-driven gap in metallic carbon nanotubes},\ }\href
  {https://doi.org/10.1103/PhysRevB.97.035445} {\bibfield  {journal} {\bibinfo
  {journal} {Phys. Rev. B}\ }\textbf {\bibinfo {volume} {97}},\ \bibinfo
  {pages} {035445} (\bibinfo {year} {2018})}\BibitemShut {NoStop}%
\bibitem [{\citenamefont {Island}\ \emph {et~al.}(2018)\citenamefont {Island},
  \citenamefont {Ostermann}, \citenamefont {Aspitarte}, \citenamefont {Minot},
  \citenamefont {Varsano}, \citenamefont {Molinari}, \citenamefont {Rontani},\
  and\ \citenamefont {Steele}}]{island2018}%
  \BibitemOpen
  \bibfield  {author} {\bibinfo {author} {\bibfnamefont {J.~O.}\ \bibnamefont
  {Island}}, \bibinfo {author} {\bibfnamefont {M.}~\bibnamefont {Ostermann}},
  \bibinfo {author} {\bibfnamefont {L.}~\bibnamefont {Aspitarte}}, \bibinfo
  {author} {\bibfnamefont {E.~D.}\ \bibnamefont {Minot}}, \bibinfo {author}
  {\bibfnamefont {D.}~\bibnamefont {Varsano}}, \bibinfo {author} {\bibfnamefont
  {E.}~\bibnamefont {Molinari}}, \bibinfo {author} {\bibfnamefont
  {M.}~\bibnamefont {Rontani}},\ and\ \bibinfo {author} {\bibfnamefont {G.~A.}\
  \bibnamefont {Steele}},\ }\bibfield  {title} {\bibinfo {title}
  {Interaction-driven giant orbital magnetic moments in carbon nanotubes},\
  }\href {https://doi.org/10.1103/PhysRevLett.121.127704} {\bibfield  {journal}
  {\bibinfo  {journal} {Phys. Rev. Lett.}\ }\textbf {\bibinfo {volume} {121}},\
  \bibinfo {pages} {127704} (\bibinfo {year} {2018})}\BibitemShut {NoStop}%
\bibitem [{\citenamefont {Feldman}\ \emph {et~al.}(2009)\citenamefont
  {Feldman}, \citenamefont {Martin},\ and\ \citenamefont
  {Yacoby}}]{Feldman2009}%
  \BibitemOpen
  \bibfield  {author} {\bibinfo {author} {\bibfnamefont {B.~E.}\ \bibnamefont
  {Feldman}}, \bibinfo {author} {\bibfnamefont {J.}~\bibnamefont {Martin}},\
  and\ \bibinfo {author} {\bibfnamefont {A.}~\bibnamefont {Yacoby}},\
  }\bibfield  {title} {\bibinfo {title} {Broken-symmetry states and divergent
  resistance in suspended bilayer graphene},\ }\href
  {https://doi.org/10.1038/NPHYS1406} {\bibfield  {journal} {\bibinfo
  {journal} {Nature Physics}\ }\textbf {\bibinfo {volume} {5}},\ \bibinfo
  {pages} {889} (\bibinfo {year} {2009})}\BibitemShut {NoStop}%
\bibitem [{\citenamefont {Freitag}\ \emph {et~al.}(2012)\citenamefont
  {Freitag}, \citenamefont {Trbovic}, \citenamefont {Weiss},\ and\
  \citenamefont {Sch{\"o}nenberger}}]{Freitag2012}%
  \BibitemOpen
  \bibfield  {author} {\bibinfo {author} {\bibfnamefont {F.}~\bibnamefont
  {Freitag}}, \bibinfo {author} {\bibfnamefont {J.}~\bibnamefont {Trbovic}},
  \bibinfo {author} {\bibfnamefont {M.}~\bibnamefont {Weiss}},\ and\ \bibinfo
  {author} {\bibfnamefont {C.}~\bibnamefont {Sch{\"o}nenberger}},\ }\bibfield
  {title} {\bibinfo {title} {Spontaneously gapped ground state in suspended
  bilayer graphene},\ }\href {https://doi.org/10.1103/PhysRevLett.108.076602}
  {\bibfield  {journal} {\bibinfo  {journal} {Phys. Rev. Lett.}\ }\textbf
  {\bibinfo {volume} {108}},\ \bibinfo {pages} {076602} (\bibinfo {year}
  {2012})}\BibitemShut {NoStop}%
\bibitem [{\citenamefont {Velasco}\ \emph {et~al.}(2012)\citenamefont
  {Velasco}, \citenamefont {Jing}, \citenamefont {Bao}, \citenamefont {Lee},
  \citenamefont {Kratz}, \citenamefont {Aji}, \citenamefont {Bockrath},
  \citenamefont {Lau}, \citenamefont {Varma}, \citenamefont {Stillwell},
  \citenamefont {Smirnov}, \citenamefont {Zhang}, \citenamefont {Jung},\ and\
  \citenamefont {MacDonald}}]{Velasco2012}%
  \BibitemOpen
  \bibfield  {author} {\bibinfo {author} {\bibfnamefont {J.}~\bibnamefont
  {Velasco}, \bibfnamefont {Jr.}}, \bibinfo {author} {\bibfnamefont
  {L.}~\bibnamefont {Jing}}, \bibinfo {author} {\bibfnamefont {W.}~\bibnamefont
  {Bao}}, \bibinfo {author} {\bibfnamefont {Y.}~\bibnamefont {Lee}}, \bibinfo
  {author} {\bibfnamefont {P.}~\bibnamefont {Kratz}}, \bibinfo {author}
  {\bibfnamefont {V.}~\bibnamefont {Aji}}, \bibinfo {author} {\bibfnamefont
  {M.}~\bibnamefont {Bockrath}}, \bibinfo {author} {\bibfnamefont {C.~N.}\
  \bibnamefont {Lau}}, \bibinfo {author} {\bibfnamefont {C.}~\bibnamefont
  {Varma}}, \bibinfo {author} {\bibfnamefont {R.}~\bibnamefont {Stillwell}},
  \bibinfo {author} {\bibfnamefont {D.}~\bibnamefont {Smirnov}}, \bibinfo
  {author} {\bibfnamefont {F.}~\bibnamefont {Zhang}}, \bibinfo {author}
  {\bibfnamefont {J.}~\bibnamefont {Jung}},\ and\ \bibinfo {author}
  {\bibfnamefont {A.~H.}\ \bibnamefont {MacDonald}},\ }\bibfield  {title}
  {\bibinfo {title} {Transport spectroscopy of symmetry-broken insulating
  states in bilayer graphene},\ }\href {https://doi.org/10.1038/NNANO.2011.251}
  {\bibfield  {journal} {\bibinfo  {journal} {Nature Nanotechnology}\ }\textbf
  {\bibinfo {volume} {7}},\ \bibinfo {pages} {156} (\bibinfo {year}
  {2012})}\BibitemShut {NoStop}%
\bibitem [{\citenamefont {Bao}\ \emph {et~al.}(2012)\citenamefont {Bao},
  \citenamefont {Velasco}, \citenamefont {Zhang}, \citenamefont {Jing},
  \citenamefont {Standley}, \citenamefont {Smirnov}, \citenamefont {Bockrath},
  \citenamefont {MacDonald},\ and\ \citenamefont {Lau}}]{Bao2012}%
  \BibitemOpen
  \bibfield  {author} {\bibinfo {author} {\bibfnamefont {W.}~\bibnamefont
  {Bao}}, \bibinfo {author} {\bibfnamefont {J.}~\bibnamefont {Velasco},
  \bibfnamefont {Jr.}}, \bibinfo {author} {\bibfnamefont {F.}~\bibnamefont
  {Zhang}}, \bibinfo {author} {\bibfnamefont {L.}~\bibnamefont {Jing}},
  \bibinfo {author} {\bibfnamefont {B.}~\bibnamefont {Standley}}, \bibinfo
  {author} {\bibfnamefont {D.}~\bibnamefont {Smirnov}}, \bibinfo {author}
  {\bibfnamefont {M.}~\bibnamefont {Bockrath}}, \bibinfo {author}
  {\bibfnamefont {A.~H.}\ \bibnamefont {MacDonald}},\ and\ \bibinfo {author}
  {\bibfnamefont {C.~N.}\ \bibnamefont {Lau}},\ }\bibfield  {title} {\bibinfo
  {title} {Evidence for a spontaneous gapped state in ultraclean bilayer
  graphene},\ }\href {https://doi.org/10.1073/pnas.1205978109} {\bibfield
  {journal} {\bibinfo  {journal} {PNAS}\ }\textbf {\bibinfo {volume} {109}},\
  \bibinfo {pages} {10802} (\bibinfo {year} {2012})}\BibitemShut {NoStop}%
\bibitem [{\citenamefont {Geisenhof}\ \emph {et~al.}(2021)\citenamefont
  {Geisenhof}, \citenamefont {Winterer}, \citenamefont {Seiler}, \citenamefont
  {Lenz}, \citenamefont {Xu}, \citenamefont {Zhang},\ and\ \citenamefont
  {Weitz}}]{Geisenhof2021}%
  \BibitemOpen
  \bibfield  {author} {\bibinfo {author} {\bibfnamefont {F.~R.}\ \bibnamefont
  {Geisenhof}}, \bibinfo {author} {\bibfnamefont {F.}~\bibnamefont {Winterer}},
  \bibinfo {author} {\bibfnamefont {A.~M.}\ \bibnamefont {Seiler}}, \bibinfo
  {author} {\bibfnamefont {J.}~\bibnamefont {Lenz}}, \bibinfo {author}
  {\bibfnamefont {T.}~\bibnamefont {Xu}}, \bibinfo {author} {\bibfnamefont
  {F.}~\bibnamefont {Zhang}},\ and\ \bibinfo {author} {\bibfnamefont {R.~T.}\
  \bibnamefont {Weitz}},\ }\bibfield  {title} {\bibinfo {title} {Quantum
  anomalous hall octet driven by orbital magnetism in bilayer graphene},\
  }\href {https://doi.org/10.1038/s41586-021-03849-w} {\bibfield  {journal}
  {\bibinfo  {journal} {Nature}\ }\textbf {\bibinfo {volume} {598}},\ \bibinfo
  {pages} {53} (\bibinfo {year} {2021})}\BibitemShut {NoStop}%
\bibitem [{\citenamefont {Bohnen}\ \emph {et~al.}(2004)\citenamefont {Bohnen},
  \citenamefont {Heid}, \citenamefont {Liu},\ and\ \citenamefont
  {Chan}}]{Bohnen2004}%
  \BibitemOpen
  \bibfield  {author} {\bibinfo {author} {\bibfnamefont {K.}~\bibnamefont
  {Bohnen}}, \bibinfo {author} {\bibfnamefont {R.}~\bibnamefont {Heid}},
  \bibinfo {author} {\bibfnamefont {H.~J.}\ \bibnamefont {Liu}},\ and\ \bibinfo
  {author} {\bibfnamefont {C.~T.}\ \bibnamefont {Chan}},\ }\bibfield  {title}
  {\bibinfo {title} {Lattice dynamics and electron-phonon interaction in (3,3)
  carbon nanotubes},\ }\href {https://doi.org/10.1103/PhysRevLett.93.245501}
  {\bibfield  {journal} {\bibinfo  {journal} {Phys. Rev. Lett.}\ }\textbf
  {\bibinfo {volume} {93}},\ \bibinfo {pages} {245501} (\bibinfo {year}
  {2004})}\BibitemShut {NoStop}%
\bibitem [{\citenamefont {Conn{\'e}table}\ \emph {et~al.}(2005)\citenamefont
  {Conn{\'e}table}, \citenamefont {Rignanese}, \citenamefont {Charlier},\ and\
  \citenamefont {Blase}}]{Connetable2005}%
  \BibitemOpen
  \bibfield  {author} {\bibinfo {author} {\bibfnamefont {D.}~\bibnamefont
  {Conn{\'e}table}}, \bibinfo {author} {\bibfnamefont {G.}~\bibnamefont
  {Rignanese}}, \bibinfo {author} {\bibfnamefont {J.}~\bibnamefont
  {Charlier}},\ and\ \bibinfo {author} {\bibfnamefont {X.}~\bibnamefont
  {Blase}},\ }\bibfield  {title} {\bibinfo {title} {Room temperature {P}eierls
  distortion in small diameter nanotubes},\ }\href
  {https://doi.org/10.1103/PhysRevLett.94.015503} {\bibfield  {journal}
  {\bibinfo  {journal} {Phys. Rev. Lett.}\ }\textbf {\bibinfo {volume} {94}},\
  \bibinfo {pages} {015503} (\bibinfo {year} {2005})}\BibitemShut {NoStop}%
\bibitem [{\citenamefont {Chen}\ \emph {et~al.}(2008)\citenamefont {Chen},
  \citenamefont {Andreev}, \citenamefont {Tsvelik},\ and\ \citenamefont
  {Orgad}}]{Chen2008}%
  \BibitemOpen
  \bibfield  {author} {\bibinfo {author} {\bibfnamefont {W.}~\bibnamefont
  {Chen}}, \bibinfo {author} {\bibfnamefont {A.~V.}\ \bibnamefont {Andreev}},
  \bibinfo {author} {\bibfnamefont {A.~M.}\ \bibnamefont {Tsvelik}},\ and\
  \bibinfo {author} {\bibfnamefont {D.}~\bibnamefont {Orgad}},\ }\bibfield
  {title} {\bibinfo {title} {Twist instability in strongly correlated carbon
  nanotubes},\ }\href {https://doi.org/10.1103/PhysRevLett.101.246802}
  {\bibfield  {journal} {\bibinfo  {journal} {Phys. Rev. Lett.}\ }\textbf
  {\bibinfo {volume} {101}},\ \bibinfo {pages} {246802} (\bibinfo {year}
  {2008})}\BibitemShut {NoStop}%
\bibitem [{\citenamefont {Dumont}\ \emph {et~al.}(2010)\citenamefont {Dumont},
  \citenamefont {Boulanger}, \citenamefont {C{\^o}t{\'e}},\ and\ \citenamefont
  {Ernzerhof}}]{dumont2010peierls}%
  \BibitemOpen
  \bibfield  {author} {\bibinfo {author} {\bibfnamefont {G.}~\bibnamefont
  {Dumont}}, \bibinfo {author} {\bibfnamefont {P.}~\bibnamefont {Boulanger}},
  \bibinfo {author} {\bibfnamefont {M.}~\bibnamefont {C{\^o}t{\'e}}},\ and\
  \bibinfo {author} {\bibfnamefont {M.}~\bibnamefont {Ernzerhof}},\ }\bibfield
  {title} {\bibinfo {title} {Peierls instability in carbon nanotubes: A
  first-principles study},\ }\href {https://doi.org/10.1103/PhysRevB.82.035419}
  {\bibfield  {journal} {\bibinfo  {journal} {Physical Review B}\ }\textbf
  {\bibinfo {volume} {82}},\ \bibinfo {pages} {035419} (\bibinfo {year}
  {2010})}\BibitemShut {NoStop}%
\bibitem [{\citenamefont {Efroni}\ \emph {et~al.}(2017)\citenamefont {Efroni},
  \citenamefont {Ilani},\ and\ \citenamefont {Berg}}]{Efroni2017}%
  \BibitemOpen
  \bibfield  {author} {\bibinfo {author} {\bibfnamefont {Y.}~\bibnamefont
  {Efroni}}, \bibinfo {author} {\bibfnamefont {S.}~\bibnamefont {Ilani}},\ and\
  \bibinfo {author} {\bibfnamefont {E.}~\bibnamefont {Berg}},\ }\bibfield
  {title} {\bibinfo {title} {Topological transitions and fractional charges
  induced by strain and a magnetic field in carbon nanotubes},\ }\href
  {https://doi.org/10.1103/PhysRevLett.119.147704} {\bibfield  {journal}
  {\bibinfo  {journal} {Phys. Rev. Lett.}\ }\textbf {\bibinfo {volume} {119}},\
  \bibinfo {pages} {147704} (\bibinfo {year} {2017})}\BibitemShut {NoStop}%
\bibitem [{\citenamefont {Voliovich}\ \emph {et~al.}(2022)\citenamefont
  {Voliovich}, \citenamefont {Rudner}, \citenamefont {Oreg},\ and\
  \citenamefont {Berg}}]{Voliovich2022}%
  \BibitemOpen
  \bibfield  {author} {\bibinfo {author} {\bibfnamefont {A.}~\bibnamefont
  {Voliovich}}, \bibinfo {author} {\bibfnamefont {M.~S.}\ \bibnamefont
  {Rudner}}, \bibinfo {author} {\bibfnamefont {Y.}~\bibnamefont {Oreg}},\ and\
  \bibinfo {author} {\bibfnamefont {E.}~\bibnamefont {Berg}},\ }\bibfield
  {title} {\bibinfo {title} {Correlated insulating states in carbon nanotubes
  controlled by magnetic field},\ }\href
  {https://doi.org/10.1103/PhysRevB.106.235141} {\bibfield  {journal} {\bibinfo
   {journal} {Phys. Rev. B}\ }\textbf {\bibinfo {volume} {106}},\ \bibinfo
  {pages} {235141} (\bibinfo {year} {2022})}\BibitemShut {NoStop}%
\bibitem [{\citenamefont {Kostyrko}(2024)}]{Kostyrko2024}%
  \BibitemOpen
  \bibfield  {author} {\bibinfo {author} {\bibfnamefont {T.}~\bibnamefont
  {Kostyrko}},\ }\bibfield  {title} {\bibinfo {title} {A {DFT}+{U} study of
  carbon nanotubes under influence of a gate voltage},\ }\href
  {https://doi.org/https://doi.org/10.1016/j.jmmm.2024.171869} {\bibfield
  {journal} {\bibinfo  {journal} {Journal of Magnetism and Magnetic Materials}\
  }\textbf {\bibinfo {volume} {593}},\ \bibinfo {pages} {171869} (\bibinfo
  {year} {2024})}\BibitemShut {NoStop}%
\bibitem [{\citenamefont {Zhang}\ \emph {et~al.}(2011)\citenamefont {Zhang},
  \citenamefont {Jung}, \citenamefont {Fiete}, \citenamefont {Niu},\ and\
  \citenamefont {MacDonald}}]{Zhang2011}%
  \BibitemOpen
  \bibfield  {author} {\bibinfo {author} {\bibfnamefont {F.}~\bibnamefont
  {Zhang}}, \bibinfo {author} {\bibfnamefont {J.}~\bibnamefont {Jung}},
  \bibinfo {author} {\bibfnamefont {G.~A.}\ \bibnamefont {Fiete}}, \bibinfo
  {author} {\bibfnamefont {Q.}~\bibnamefont {Niu}},\ and\ \bibinfo {author}
  {\bibfnamefont {A.~H.}\ \bibnamefont {MacDonald}},\ }\bibfield  {title}
  {\bibinfo {title} {Spontaneous quantum hall states in chirally stacked
  few-layer graphene systems},\ }\href
  {https://doi.org/10.1103/PhysRevLett.106.156801} {\bibfield  {journal}
  {\bibinfo  {journal} {Phys. Rev. Lett.}\ }\textbf {\bibinfo {volume} {106}},\
  \bibinfo {pages} {156801} (\bibinfo {year} {2011})}\BibitemShut {NoStop}%
\bibitem [{\citenamefont {Maultzsch}\ \emph {et~al.}(2005)\citenamefont
  {Maultzsch}, \citenamefont {Pomraenke}, \citenamefont {Reich}, \citenamefont
  {Chang}, \citenamefont {Prezzi}, \citenamefont {Ruini}, \citenamefont
  {Molinari}, \citenamefont {Strano}, \citenamefont {Thomsen},\ and\
  \citenamefont {Lienau}}]{Maultzsch2005}%
  \BibitemOpen
  \bibfield  {author} {\bibinfo {author} {\bibfnamefont {J.}~\bibnamefont
  {Maultzsch}}, \bibinfo {author} {\bibfnamefont {R.}~\bibnamefont
  {Pomraenke}}, \bibinfo {author} {\bibfnamefont {S.}~\bibnamefont {Reich}},
  \bibinfo {author} {\bibfnamefont {E.}~\bibnamefont {Chang}}, \bibinfo
  {author} {\bibfnamefont {D.}~\bibnamefont {Prezzi}}, \bibinfo {author}
  {\bibfnamefont {A.}~\bibnamefont {Ruini}}, \bibinfo {author} {\bibfnamefont
  {E.}~\bibnamefont {Molinari}}, \bibinfo {author} {\bibfnamefont {M.~S.}\
  \bibnamefont {Strano}}, \bibinfo {author} {\bibfnamefont {C.}~\bibnamefont
  {Thomsen}},\ and\ \bibinfo {author} {\bibfnamefont {C.}~\bibnamefont
  {Lienau}},\ }\bibfield  {title} {\bibinfo {title} {Exciton binding energies
  in carbon nanotubes from two-photon photoluminescence},\ }\href
  {https://doi.org/10.1103/PhysRevB.72.241402} {\bibfield  {journal} {\bibinfo
  {journal} {Phys. Rev. B}\ }\textbf {\bibinfo {volume} {72}},\ \bibinfo
  {pages} {241402(R)} (\bibinfo {year} {2005})}\BibitemShut {NoStop}%
\bibitem [{\citenamefont {Wang}\ \emph {et~al.}(2005)\citenamefont {Wang},
  \citenamefont {Dukovic}, \citenamefont {Brus},\ and\ \citenamefont
  {Heinz}}]{Wang2005}%
  \BibitemOpen
  \bibfield  {author} {\bibinfo {author} {\bibfnamefont {F.}~\bibnamefont
  {Wang}}, \bibinfo {author} {\bibfnamefont {G.}~\bibnamefont {Dukovic}},
  \bibinfo {author} {\bibfnamefont {L.~E.}\ \bibnamefont {Brus}},\ and\
  \bibinfo {author} {\bibfnamefont {T.}~\bibnamefont {Heinz}},\ }\bibfield
  {title} {\bibinfo {title} {The optical resonances in carbon nanotubes arise
  from excitons},\ }\href {https://doi.org/10.1126/science.1110265} {\bibfield
  {journal} {\bibinfo  {journal} {Science}\ }\textbf {\bibinfo {volume}
  {308}},\ \bibinfo {pages} {838} (\bibinfo {year} {2005})}\BibitemShut
  {NoStop}%
\bibitem [{\citenamefont {Kravets}\ \emph {et~al.}(2010)\citenamefont
  {Kravets}, \citenamefont {Grigorenko}, \citenamefont {Nair}, \citenamefont
  {Blake}, \citenamefont {Anissimova}, \citenamefont {Novoselov},\ and\
  \citenamefont {Geim}}]{Kravets2010}%
  \BibitemOpen
  \bibfield  {author} {\bibinfo {author} {\bibfnamefont {V.~G.}\ \bibnamefont
  {Kravets}}, \bibinfo {author} {\bibfnamefont {A.~N.}\ \bibnamefont
  {Grigorenko}}, \bibinfo {author} {\bibfnamefont {R.~R.}\ \bibnamefont
  {Nair}}, \bibinfo {author} {\bibfnamefont {P.}~\bibnamefont {Blake}},
  \bibinfo {author} {\bibfnamefont {S.}~\bibnamefont {Anissimova}}, \bibinfo
  {author} {\bibfnamefont {K.~S.}\ \bibnamefont {Novoselov}},\ and\ \bibinfo
  {author} {\bibfnamefont {A.~K.}\ \bibnamefont {Geim}},\ }\bibfield  {title}
  {\bibinfo {title} {Spectroscopic ellipsometry of graphene and an
  exciton-shifted van {H}ove peak in absorption},\ }\href
  {https://doi.org/10.1103/PhysRevB.81.155413} {\bibfield  {journal} {\bibinfo
  {journal} {Phys. Rev. B}\ }\textbf {\bibinfo {volume} {81}},\ \bibinfo
  {pages} {155413} (\bibinfo {year} {2010})}\BibitemShut {NoStop}%
\bibitem [{\citenamefont {Ju}\ \emph {et~al.}(2017)\citenamefont {Ju},
  \citenamefont {Wang}, \citenamefont {Cao}, \citenamefont {Taniguchi},
  \citenamefont {Watanabe}, \citenamefont {Louie}, \citenamefont {Rana},
  \citenamefont {Park}, \citenamefont {Hone}, \citenamefont {Wang},\ and\
  \citenamefont {McEuen}}]{Ju2017}%
  \BibitemOpen
  \bibfield  {author} {\bibinfo {author} {\bibfnamefont {L.}~\bibnamefont
  {Ju}}, \bibinfo {author} {\bibfnamefont {L.}~\bibnamefont {Wang}}, \bibinfo
  {author} {\bibfnamefont {T.}~\bibnamefont {Cao}}, \bibinfo {author}
  {\bibfnamefont {T.}~\bibnamefont {Taniguchi}}, \bibinfo {author}
  {\bibfnamefont {K.}~\bibnamefont {Watanabe}}, \bibinfo {author}
  {\bibfnamefont {S.~G.}\ \bibnamefont {Louie}}, \bibinfo {author}
  {\bibfnamefont {F.}~\bibnamefont {Rana}}, \bibinfo {author} {\bibfnamefont
  {J.}~\bibnamefont {Park}}, \bibinfo {author} {\bibfnamefont {J.}~\bibnamefont
  {Hone}}, \bibinfo {author} {\bibfnamefont {F.}~\bibnamefont {Wang}},\ and\
  \bibinfo {author} {\bibfnamefont {P.~L.}\ \bibnamefont {McEuen}},\ }\bibfield
   {title} {\bibinfo {title} {Tunable excitons in bilayer graphene},\ }\href
  {https://doi.org/10.1126/science.aam9175} {\bibfield  {journal} {\bibinfo
  {journal} {Science}\ }\textbf {\bibinfo {volume} {358}},\ \bibinfo {pages}
  {907} (\bibinfo {year} {2017})}\BibitemShut {NoStop}%
\bibitem [{\citenamefont {Wang}\ \emph {et~al.}(2007)\citenamefont {Wang},
  \citenamefont {Cho}, \citenamefont {Kessler}, \citenamefont {Deslippe},
  \citenamefont {Schuck}, \citenamefont {Louie}, \citenamefont {Zettl},
  \citenamefont {Heinz},\ and\ \citenamefont {Shen}}]{Wang2007}%
  \BibitemOpen
  \bibfield  {author} {\bibinfo {author} {\bibfnamefont {F.}~\bibnamefont
  {Wang}}, \bibinfo {author} {\bibfnamefont {D.~J.}\ \bibnamefont {Cho}},
  \bibinfo {author} {\bibfnamefont {B.}~\bibnamefont {Kessler}}, \bibinfo
  {author} {\bibfnamefont {J.}~\bibnamefont {Deslippe}}, \bibinfo {author}
  {\bibfnamefont {P.~J.}\ \bibnamefont {Schuck}}, \bibinfo {author}
  {\bibfnamefont {S.~G.}\ \bibnamefont {Louie}}, \bibinfo {author}
  {\bibfnamefont {A.}~\bibnamefont {Zettl}}, \bibinfo {author} {\bibfnamefont
  {T.~F.}\ \bibnamefont {Heinz}},\ and\ \bibinfo {author} {\bibfnamefont
  {Y.~R.}\ \bibnamefont {Shen}},\ }\bibfield  {title} {\bibinfo {title}
  {Observation of excitons in one-dimensional metallic single-walled carbon
  nanotubes},\ }\href {https://doi.org/10.1103/PhysRevLett.99.227401}
  {\bibfield  {journal} {\bibinfo  {journal} {Phys. Rev. Lett.}\ }\textbf
  {\bibinfo {volume} {99}},\ \bibinfo {pages} {227401} (\bibinfo {year}
  {2007})}\BibitemShut {NoStop}%
\bibitem [{\citenamefont {Khveshchenko}(2001)}]{Khveshchenko2001}%
  \BibitemOpen
  \bibfield  {author} {\bibinfo {author} {\bibfnamefont {D.~V.}\ \bibnamefont
  {Khveshchenko}},\ }\bibfield  {title} {\bibinfo {title} {Ghost excitonic
  insulator transition in layered graphite},\ }\href
  {https://doi.org/10.1103/PhysRevLett.87.246802} {\bibfield  {journal}
  {\bibinfo  {journal} {Phys. Rev. Lett.}\ }\textbf {\bibinfo {volume} {87}},\
  \bibinfo {pages} {246802} (\bibinfo {year} {2001})}\BibitemShut {NoStop}%
\bibitem [{\citenamefont {Varsano}\ \emph {et~al.}(2017)\citenamefont
  {Varsano}, \citenamefont {Sorella}, \citenamefont {Sangalli}, \citenamefont
  {Barborini}, \citenamefont {Corni}, \citenamefont {Molinari},\ and\
  \citenamefont {Rontani}}]{varsano2017carbon}%
  \BibitemOpen
  \bibfield  {author} {\bibinfo {author} {\bibfnamefont {D.}~\bibnamefont
  {Varsano}}, \bibinfo {author} {\bibfnamefont {S.}~\bibnamefont {Sorella}},
  \bibinfo {author} {\bibfnamefont {D.}~\bibnamefont {Sangalli}}, \bibinfo
  {author} {\bibfnamefont {M.}~\bibnamefont {Barborini}}, \bibinfo {author}
  {\bibfnamefont {S.}~\bibnamefont {Corni}}, \bibinfo {author} {\bibfnamefont
  {E.}~\bibnamefont {Molinari}},\ and\ \bibinfo {author} {\bibfnamefont
  {M.}~\bibnamefont {Rontani}},\ }\bibfield  {title} {\bibinfo {title} {Carbon
  nanotubes as excitonic insulators},\ }\href
  {https://doi.org/10.1038/s41467-017-01660-8} {\bibfield  {journal} {\bibinfo
  {journal} {Nature Communications}\ }\textbf {\bibinfo {volume} {8}},\
  \bibinfo {pages} {1461} (\bibinfo {year} {2017})}\BibitemShut {NoStop}%
\bibitem [{\citenamefont {J{\`e}rome}\ \emph {et~al.}(1967)\citenamefont
  {J{\`e}rome}, \citenamefont {Rice},\ and\ \citenamefont {Kohn}}]{Kohn1967}%
  \BibitemOpen
  \bibfield  {author} {\bibinfo {author} {\bibfnamefont {D.}~\bibnamefont
  {J{\`e}rome}}, \bibinfo {author} {\bibfnamefont {T.~M.}\ \bibnamefont
  {Rice}},\ and\ \bibinfo {author} {\bibfnamefont {W.}~\bibnamefont {Kohn}},\
  }\bibfield  {title} {\bibinfo {title} {Excitonic insulator},\ }\href
  {https://doi.org/10.1103/PhysRev.158.462} {\bibfield  {journal} {\bibinfo
  {journal} {Phys. Rev.}\ }\textbf {\bibinfo {volume} {158}},\ \bibinfo {pages}
  {462} (\bibinfo {year} {1967})}\BibitemShut {NoStop}%
\bibitem [{\citenamefont {Halperin}\ and\ \citenamefont
  {Rice}(1968)}]{Halperin1968}%
  \BibitemOpen
  \bibfield  {author} {\bibinfo {author} {\bibfnamefont {B.~I.}\ \bibnamefont
  {Halperin}}\ and\ \bibinfo {author} {\bibfnamefont {T.~M.}\ \bibnamefont
  {Rice}},\ }\bibfield  {title} {\bibinfo {title} {The excitonic state at the
  semiconductor-semimetal transition},\ }\href
  {https://doi.org/10.1016/S0081-1947(08)60740-7} {\bibfield  {journal}
  {\bibinfo  {journal} {Solid State Phys.}\ }\textbf {\bibinfo {volume} {21}},\
  \bibinfo {pages} {115} (\bibinfo {year} {1968})}\BibitemShut {NoStop}%
\bibitem [{\citenamefont {Hellgren}\ \emph {et~al.}(2018)\citenamefont
  {Hellgren}, \citenamefont {Baima},\ and\ \citenamefont
  {Acheche}}]{Hellgren2018}%
  \BibitemOpen
  \bibfield  {author} {\bibinfo {author} {\bibfnamefont {M.}~\bibnamefont
  {Hellgren}}, \bibinfo {author} {\bibfnamefont {J.}~\bibnamefont {Baima}},\
  and\ \bibinfo {author} {\bibfnamefont {A.}~\bibnamefont {Acheche}},\
  }\bibfield  {title} {\bibinfo {title} {Exciton {P}eierls mechanism and
  universal many-body gaps in carbon nanotubes},\ }\href
  {https://doi.org/10.1103/PhysRevB.98.201103} {\bibfield  {journal} {\bibinfo
  {journal} {Phys. Rev. B}\ }\textbf {\bibinfo {volume} {98}},\ \bibinfo
  {pages} {201103} (\bibinfo {year} {2018})}\BibitemShut {NoStop}%
\bibitem [{\citenamefont {Okamoto}\ \emph {et~al.}(2018)\citenamefont
  {Okamoto}, \citenamefont {Mathey},\ and\ \citenamefont
  {Huang}}]{Okamoto2018}%
  \BibitemOpen
  \bibfield  {author} {\bibinfo {author} {\bibfnamefont {J.}~\bibnamefont
  {Okamoto}}, \bibinfo {author} {\bibfnamefont {L.}~\bibnamefont {Mathey}},\
  and\ \bibinfo {author} {\bibfnamefont {W.-M.}\ \bibnamefont {Huang}},\
  }\bibfield  {title} {\bibinfo {title} {Influence of electron-phonon coupling
  on the low-temperature phases of metallic single-wall carbon nanotubes},\
  }\href {https://doi.org/10.1103/PhysRevB.98.205122} {\bibfield  {journal}
  {\bibinfo  {journal} {Phys. Rev. B}\ }\textbf {\bibinfo {volume} {98}},\
  \bibinfo {pages} {205122} (\bibinfo {year} {2018})}\BibitemShut {NoStop}%
\bibitem [{\citenamefont {Barborini}\ \emph {et~al.}(2022)\citenamefont
  {Barborini}, \citenamefont {Calandra}, \citenamefont {Mauri}, \citenamefont
  {Wirtz},\ and\ \citenamefont {Cudazzo}}]{Barborini2022}%
  \BibitemOpen
  \bibfield  {author} {\bibinfo {author} {\bibfnamefont {M.}~\bibnamefont
  {Barborini}}, \bibinfo {author} {\bibfnamefont {M.}~\bibnamefont {Calandra}},
  \bibinfo {author} {\bibfnamefont {F.}~\bibnamefont {Mauri}}, \bibinfo
  {author} {\bibfnamefont {L.}~\bibnamefont {Wirtz}},\ and\ \bibinfo {author}
  {\bibfnamefont {P.}~\bibnamefont {Cudazzo}},\ }\bibfield  {title} {\bibinfo
  {title} {Excitonic-insulator instability and {P}eierls distortion in
  one-dimensional semimetals},\ }\href
  {https://doi.org/10.1103/PhysRevB.105.075122} {\bibfield  {journal} {\bibinfo
   {journal} {Phys. Rev. B}\ }\textbf {\bibinfo {volume} {105}},\ \bibinfo
  {pages} {075122} (\bibinfo {year} {2022})}\BibitemShut {NoStop}%
\bibitem [{\citenamefont {Elliott}\ \emph {et~al.}(2004)\citenamefont
  {Elliott}, \citenamefont {Sandler}, \citenamefont {Windle}, \citenamefont
  {Young},\ and\ \citenamefont {Shaffer}}]{elliott2004}%
  \BibitemOpen
  \bibfield  {author} {\bibinfo {author} {\bibfnamefont {J.~A.}\ \bibnamefont
  {Elliott}}, \bibinfo {author} {\bibfnamefont {J.~K.~W.}\ \bibnamefont
  {Sandler}}, \bibinfo {author} {\bibfnamefont {A.~H.}\ \bibnamefont {Windle}},
  \bibinfo {author} {\bibfnamefont {R.~J.}\ \bibnamefont {Young}},\ and\
  \bibinfo {author} {\bibfnamefont {M.~S.~P.}\ \bibnamefont {Shaffer}},\
  }\bibfield  {title} {\bibinfo {title} {Collapse of single-wall carbon
  nanotubes is diameter dependent},\ }\href
  {https://doi.org/10.1103/PhysRevLett.92.095501} {\bibfield  {journal}
  {\bibinfo  {journal} {Phys. Rev. Lett.}\ }\textbf {\bibinfo {volume} {92}},\
  \bibinfo {pages} {095501} (\bibinfo {year} {2004})}\BibitemShut {NoStop}%
\bibitem [{\citenamefont {Sesti}\ \emph {et~al.}(2022)\citenamefont {Sesti},
  \citenamefont {Varsano}, \citenamefont {Molinari},\ and\ \citenamefont
  {Rontani}}]{Sesti2022}%
  \BibitemOpen
  \bibfield  {author} {\bibinfo {author} {\bibfnamefont {G.}~\bibnamefont
  {Sesti}}, \bibinfo {author} {\bibfnamefont {D.}~\bibnamefont {Varsano}},
  \bibinfo {author} {\bibfnamefont {E.}~\bibnamefont {Molinari}},\ and\
  \bibinfo {author} {\bibfnamefont {M.}~\bibnamefont {Rontani}},\ }\bibfield
  {title} {\bibinfo {title} {Anomalous screening in narrow-gap carbon
  nanotubes},\ }\href {https://doi.org/10.1103/PhysRevB.105.195404} {\bibfield
  {journal} {\bibinfo  {journal} {Phys. Rev. B}\ }\textbf {\bibinfo {volume}
  {105}},\ \bibinfo {pages} {195404} (\bibinfo {year} {2022})}\BibitemShut
  {NoStop}%
\bibitem [{\citenamefont {Ando}(1997)}]{Ando1997}%
  \BibitemOpen
  \bibfield  {author} {\bibinfo {author} {\bibfnamefont {T.}~\bibnamefont
  {Ando}},\ }\bibfield  {title} {\bibinfo {title} {Excitons in carbon
  nanotubes},\ }\href {https://doi.org/10.1143/JPSJ.66.1066} {\bibfield
  {journal} {\bibinfo  {journal} {J. Phys. Soc. Jpn.}\ }\textbf {\bibinfo
  {volume} {66}},\ \bibinfo {pages} {1066} (\bibinfo {year}
  {1997})}\BibitemShut {NoStop}%
\bibitem [{\citenamefont {Spataru}\ \emph {et~al.}(2004)\citenamefont
  {Spataru}, \citenamefont {Ismail-Beigi}, \citenamefont {Benedict},\ and\
  \citenamefont {Louie}}]{Spataru2004}%
  \BibitemOpen
  \bibfield  {author} {\bibinfo {author} {\bibfnamefont {C.~D.}\ \bibnamefont
  {Spataru}}, \bibinfo {author} {\bibfnamefont {S.}~\bibnamefont
  {Ismail-Beigi}}, \bibinfo {author} {\bibfnamefont {L.~X.}\ \bibnamefont
  {Benedict}},\ and\ \bibinfo {author} {\bibfnamefont {S.~G.}\ \bibnamefont
  {Louie}},\ }\bibfield  {title} {\bibinfo {title} {Excitonic effects and
  optical spectra of single-walled carbon nanotubes},\ }\href
  {https://doi.org/10.1103/PhysRevLett.92.077402} {\bibfield  {journal}
  {\bibinfo  {journal} {Phys. Rev. Lett.}\ }\textbf {\bibinfo {volume} {92}},\
  \bibinfo {pages} {077402} (\bibinfo {year} {2004})}\BibitemShut {NoStop}%
\bibitem [{\citenamefont {Kane}\ and\ \citenamefont
  {Mele}(1997)}]{kane1997size}%
  \BibitemOpen
  \bibfield  {author} {\bibinfo {author} {\bibfnamefont {C.~L.}\ \bibnamefont
  {Kane}}\ and\ \bibinfo {author} {\bibfnamefont {E.}~\bibnamefont {Mele}},\
  }\bibfield  {title} {\bibinfo {title} {Size, shape, and low energy electronic
  structure of carbon nanotubes},\ }\href
  {https://doi.org/10.1103/PhysRevLett.78.1932} {\bibfield  {journal} {\bibinfo
   {journal} {Physical Review Letters}\ }\textbf {\bibinfo {volume} {78}},\
  \bibinfo {pages} {1932} (\bibinfo {year} {1997})}\BibitemShut {NoStop}%
\bibitem [{\citenamefont {Ajiki}\ and\ \citenamefont
  {Ando}(1993)}]{ajiki1993electronic}%
  \BibitemOpen
  \bibfield  {author} {\bibinfo {author} {\bibfnamefont {H.}~\bibnamefont
  {Ajiki}}\ and\ \bibinfo {author} {\bibfnamefont {T.}~\bibnamefont {Ando}},\
  }\bibfield  {title} {\bibinfo {title} {Electronic states of carbon
  nanotubes},\ }\href {https://doi.org/10.1143/JPSJ.62.1255} {\bibfield
  {journal} {\bibinfo  {journal} {Journal of the Physical Society of Japan}\
  }\textbf {\bibinfo {volume} {62}},\ \bibinfo {pages} {1255} (\bibinfo {year}
  {1993})}\BibitemShut {NoStop}%
\bibitem [{\citenamefont {Ando}(2006)}]{Ando2006}%
  \BibitemOpen
  \bibfield  {author} {\bibinfo {author} {\bibfnamefont {T.}~\bibnamefont
  {Ando}},\ }\bibfield  {title} {\bibinfo {title} {Effects of valley mixing and
  exchange on excitons in carbon nanotubes with {A}haronov-{B}ohm flux},\
  }\href {https://doi.org/10.1143/JPSJ.75.024707} {\bibfield  {journal}
  {\bibinfo  {journal} {J. Phys. Soc. Jpn.}\ }\textbf {\bibinfo {volume}
  {75}},\ \bibinfo {pages} {024707} (\bibinfo {year} {2006})}\BibitemShut
  {NoStop}%
\bibitem [{\citenamefont {Keldysh}(1995)}]{Keldysh1995}%
  \BibitemOpen
  \bibfield  {author} {\bibinfo {author} {\bibfnamefont {L.~V.}\ \bibnamefont
  {Keldysh}},\ }\bibfield  {title} {\bibinfo {title} {Macroscopic coherent
  states of excitons in semiconductors},\ }in\ \href@noop {} {\emph {\bibinfo
  {booktitle} {Bose-{E}instein condensation}}},\ \bibinfo {editor} {edited by\
  \bibinfo {editor} {\bibfnamefont {A.}~\bibnamefont {Griffin}}, \bibinfo
  {editor} {\bibfnamefont {D.~W.}\ \bibnamefont {Snoke}},\ and\ \bibinfo
  {editor} {\bibfnamefont {S.}~\bibnamefont {Stringari}}}\ (\bibinfo
  {publisher} {Cambridge University Press},\ \bibinfo {address} {Cambridge,
  UK},\ \bibinfo {year} {1995})\ Chap.~\bibinfo {chapter} {12}, pp.\ \bibinfo
  {pages} {246--280}\BibitemShut {NoStop}%
\bibitem [{\citenamefont {Khivrich}\ \emph {et~al.}(2019)\citenamefont
  {Khivrich}, \citenamefont {Clerk},\ and\ \citenamefont
  {Ilani}}]{Khivrich2019}%
  \BibitemOpen
  \bibfield  {author} {\bibinfo {author} {\bibfnamefont {I.}~\bibnamefont
  {Khivrich}}, \bibinfo {author} {\bibfnamefont {A.~A.}\ \bibnamefont
  {Clerk}},\ and\ \bibinfo {author} {\bibfnamefont {S.}~\bibnamefont {Ilani}},\
  }\bibfield  {title} {\bibinfo {title} {Nanomechanical pump-probe measurements
  of insulating electronic states in a carbon nanotube},\ }\href
  {https://doi.org/10,1038/s41565-018-0341-6} {\bibfield  {journal} {\bibinfo
  {journal} {Nature Nanotech.}\ }\textbf {\bibinfo {volume} {14}},\ \bibinfo
  {pages} {161} (\bibinfo {year} {2019})}\BibitemShut {NoStop}%
\bibitem [{\citenamefont {Neto}(2009)}]{CastroNeto2009b}%
  \BibitemOpen
  \bibfield  {author} {\bibinfo {author} {\bibfnamefont {A.~H.~C.}\
  \bibnamefont {Neto}},\ }\bibfield  {title} {\bibinfo {title} {Pauling's
  dreams for graphene},\ }\href@noop {} {\bibfield  {journal} {\bibinfo
  {journal} {Physics}\ }\textbf {\bibinfo {volume} {2}},\ \bibinfo {pages} {30}
  (\bibinfo {year} {2009})}\BibitemShut {NoStop}%
\bibitem [{\citenamefont {Drut}\ and\ \citenamefont
  {L{\"a}nde}(2009)}]{Drut2009}%
  \BibitemOpen
  \bibfield  {author} {\bibinfo {author} {\bibfnamefont {J.~E.}\ \bibnamefont
  {Drut}}\ and\ \bibinfo {author} {\bibfnamefont {T.~A.}\ \bibnamefont
  {L{\"a}nde}},\ }\bibfield  {title} {\bibinfo {title} {Is graphene in vacuum
  an insulator?},\ }\href@noop {} {\bibfield  {journal} {\bibinfo  {journal}
  {Phys. Rev. Lett.}\ }\textbf {\bibinfo {volume} {102}},\ \bibinfo {pages}
  {026802} (\bibinfo {year} {2009})}\BibitemShut {NoStop}%
\bibitem [{\citenamefont {Gamayun}\ \emph {et~al.}(2009)\citenamefont
  {Gamayun}, \citenamefont {Gorbar},\ and\ \citenamefont
  {Gusynin}}]{Gamayun2009}%
  \BibitemOpen
  \bibfield  {author} {\bibinfo {author} {\bibfnamefont {O.~V.}\ \bibnamefont
  {Gamayun}}, \bibinfo {author} {\bibfnamefont {E.~V.}\ \bibnamefont
  {Gorbar}},\ and\ \bibinfo {author} {\bibfnamefont {V.~P.}\ \bibnamefont
  {Gusynin}},\ }\bibfield  {title} {\bibinfo {title} {Supercritical {C}oulomb
  center and excitonic instability in graphene},\ }\href
  {https://doi.org/10.1103/PhysRevB.80.165429} {\bibfield  {journal} {\bibinfo
  {journal} {Phys. Rev. B}\ }\textbf {\bibinfo {volume} {80}},\ \bibinfo
  {pages} {165429} (\bibinfo {year} {2009})}\BibitemShut {NoStop}%
\bibitem [{\citenamefont {Sabio}\ \emph {et~al.}(2010)\citenamefont {Sabio},
  \citenamefont {Sols},\ and\ \citenamefont {Guinea}}]{Sabio2010}%
  \BibitemOpen
  \bibfield  {author} {\bibinfo {author} {\bibfnamefont {J.}~\bibnamefont
  {Sabio}}, \bibinfo {author} {\bibfnamefont {F.}~\bibnamefont {Sols}},\ and\
  \bibinfo {author} {\bibfnamefont {F.}~\bibnamefont {Guinea}},\ }\bibfield
  {title} {\bibinfo {title} {Two-body problem in graphene},\ }\href
  {https://doi.org/10.1103/PhysRevB.81.045428} {\bibfield  {journal} {\bibinfo
  {journal} {Phys. Rev. B}\ }\textbf {\bibinfo {volume} {81}},\ \bibinfo
  {pages} {045428} (\bibinfo {year} {2010})}\BibitemShut {NoStop}%
\bibitem [{\citenamefont {Gr{\"o}nqvist}\ \emph {et~al.}(2012)\citenamefont
  {Gr{\"o}nqvist}, \citenamefont {Stroucken}, \citenamefont {Lindberg},\ and\
  \citenamefont {Koch}}]{Gronqvist2012}%
  \BibitemOpen
  \bibfield  {author} {\bibinfo {author} {\bibfnamefont {J.~H.}\ \bibnamefont
  {Gr{\"o}nqvist}}, \bibinfo {author} {\bibfnamefont {T.}~\bibnamefont
  {Stroucken}}, \bibinfo {author} {\bibfnamefont {M.}~\bibnamefont
  {Lindberg}},\ and\ \bibinfo {author} {\bibfnamefont {S.~W.}\ \bibnamefont
  {Koch}},\ }\bibfield  {title} {\bibinfo {title} {Wannier excitons signalling
  strong {C}oulomb coupling in graphene},\ }\href@noop {} {\bibfield  {journal}
  {\bibinfo  {journal} {Eur. Phys. J. B}\ }\textbf {\bibinfo {volume} {85}},\
  \bibinfo {pages} {395} (\bibinfo {year} {2012})}\BibitemShut {NoStop}%
\bibitem [{\citenamefont {Berestetskii}\ \emph {et~al.}(1982)\citenamefont
  {Berestetskii}, \citenamefont {Lifshitz},\ and\ \citenamefont
  {Pitaevskii}}]{Berestetskii1982}%
  \BibitemOpen
  \bibfield  {author} {\bibinfo {author} {\bibfnamefont {V.~B.}\ \bibnamefont
  {Berestetskii}}, \bibinfo {author} {\bibfnamefont {E.~M.}\ \bibnamefont
  {Lifshitz}},\ and\ \bibinfo {author} {\bibfnamefont {L.~P.}\ \bibnamefont
  {Pitaevskii}},\ }\href@noop {} {\emph {\bibinfo {title} {Quantum
  Electrodynamics}}}\ (\bibinfo  {publisher} {Pergamon Press},\ \bibinfo
  {address} {Oxford},\ \bibinfo {year} {1982})\BibitemShut {NoStop}%
\bibitem [{\citenamefont {Min}\ \emph {et~al.}(2008{\natexlab{a}})\citenamefont
  {Min}, \citenamefont {Borghi}, \citenamefont {Polini},\ and\ \citenamefont
  {MacDonald}}]{Min2008b}%
  \BibitemOpen
  \bibfield  {author} {\bibinfo {author} {\bibfnamefont {H.}~\bibnamefont
  {Min}}, \bibinfo {author} {\bibfnamefont {G.}~\bibnamefont {Borghi}},
  \bibinfo {author} {\bibfnamefont {M.}~\bibnamefont {Polini}},\ and\ \bibinfo
  {author} {\bibfnamefont {A.~H.}\ \bibnamefont {MacDonald}},\ }\bibfield
  {title} {\bibinfo {title} {Pseudospin magnetism in graphene},\ }\href@noop {}
  {\bibfield  {journal} {\bibinfo  {journal} {Phys. Rev. B}\ }\textbf {\bibinfo
  {volume} {77}},\ \bibinfo {pages} {041407(R)} (\bibinfo {year}
  {2008}{\natexlab{a}})}\BibitemShut {NoStop}%
\bibitem [{\citenamefont {Min}\ \emph {et~al.}(2008{\natexlab{b}})\citenamefont
  {Min}, \citenamefont {Bistritzer}, \citenamefont {Su},\ and\ \citenamefont
  {MacDonald}}]{Min2008}%
  \BibitemOpen
  \bibfield  {author} {\bibinfo {author} {\bibfnamefont {H.}~\bibnamefont
  {Min}}, \bibinfo {author} {\bibfnamefont {R.}~\bibnamefont {Bistritzer}},
  \bibinfo {author} {\bibfnamefont {J.}~\bibnamefont {Su}},\ and\ \bibinfo
  {author} {\bibfnamefont {A.~H.}\ \bibnamefont {MacDonald}},\ }\bibfield
  {title} {\bibinfo {title} {Room-temperature superfluidity in graphene
  bilayers},\ }\href@noop {} {\bibfield  {journal} {\bibinfo  {journal} {Phys.
  Rev. B}\ }\textbf {\bibinfo {volume} {78}},\ \bibinfo {pages} {121401(R)}
  (\bibinfo {year} {2008}{\natexlab{b}})}\BibitemShut {NoStop}%
\bibitem [{\citenamefont {Zhang}\ and\ \citenamefont
  {Joglekar}(2008)}]{Zhang2008}%
  \BibitemOpen
  \bibfield  {author} {\bibinfo {author} {\bibfnamefont {C.}~\bibnamefont
  {Zhang}}\ and\ \bibinfo {author} {\bibfnamefont {Y.~N.}\ \bibnamefont
  {Joglekar}},\ }\bibfield  {title} {\bibinfo {title} {Excitonic condensation
  of massless fermions in graphene bilayers},\ }\href@noop {} {\bibfield
  {journal} {\bibinfo  {journal} {Phys. Rev. B}\ }\textbf {\bibinfo {volume}
  {77}},\ \bibinfo {pages} {233405} (\bibinfo {year} {2008})}\BibitemShut
  {NoStop}%
\bibitem [{\citenamefont {Kharitonov}\ and\ \citenamefont
  {Efetov}(2008)}]{Kharitonov2008}%
  \BibitemOpen
  \bibfield  {author} {\bibinfo {author} {\bibfnamefont {M.~Y.}\ \bibnamefont
  {Kharitonov}}\ and\ \bibinfo {author} {\bibfnamefont {K.~B.}\ \bibnamefont
  {Efetov}},\ }\bibfield  {title} {\bibinfo {title} {Electron screening and
  excitonic condensation in double-layer graphene systems},\ }\href@noop {}
  {\bibfield  {journal} {\bibinfo  {journal} {Phys. Rev. B}\ }\textbf {\bibinfo
  {volume} {78}},\ \bibinfo {pages} {241401(R)} (\bibinfo {year}
  {2008})}\BibitemShut {NoStop}%
\bibitem [{\citenamefont {Nandkishore}\ and\ \citenamefont
  {Levitov}(2010)}]{Nandkishore2010}%
  \BibitemOpen
  \bibfield  {author} {\bibinfo {author} {\bibfnamefont {R.}~\bibnamefont
  {Nandkishore}}\ and\ \bibinfo {author} {\bibfnamefont {L.}~\bibnamefont
  {Levitov}},\ }\bibfield  {title} {\bibinfo {title} {Dynamical screening and
  excitonic instability in bilayer graphene},\ }\href@noop {} {\bibfield
  {journal} {\bibinfo  {journal} {Phys. Rev. Lett.}\ }\textbf {\bibinfo
  {volume} {104}},\ \bibinfo {pages} {156803} (\bibinfo {year}
  {2010})}\BibitemShut {NoStop}%
\bibitem [{\citenamefont {Rozzi}\ \emph {et~al.}(2006)\citenamefont {Rozzi},
  \citenamefont {Varsano}, \citenamefont {Marini}, \citenamefont {Gross},\ and\
  \citenamefont {Rubio}}]{rozzi2006exact}%
  \BibitemOpen
  \bibfield  {author} {\bibinfo {author} {\bibfnamefont {C.~A.}\ \bibnamefont
  {Rozzi}}, \bibinfo {author} {\bibfnamefont {D.}~\bibnamefont {Varsano}},
  \bibinfo {author} {\bibfnamefont {A.}~\bibnamefont {Marini}}, \bibinfo
  {author} {\bibfnamefont {E.~K.~U.}\ \bibnamefont {Gross}},\ and\ \bibinfo
  {author} {\bibfnamefont {A.}~\bibnamefont {Rubio}},\ }\bibfield  {title}
  {\bibinfo {title} {Exact coulomb cutoff technique for supercell
  calculations},\ }\href {https://doi.org/10.1103/PhysRevB.73.205119}
  {\bibfield  {journal} {\bibinfo  {journal} {Phys. Rev. B}\ }\textbf {\bibinfo
  {volume} {73}},\ \bibinfo {pages} {205119} (\bibinfo {year}
  {2006})}\BibitemShut {NoStop}%
\bibitem [{\citenamefont {Strinati}\ \emph {et~al.}(1980)\citenamefont
  {Strinati}, \citenamefont {Mattausch},\ and\ \citenamefont
  {Hanke}}]{strinati1980dynamical}%
  \BibitemOpen
  \bibfield  {author} {\bibinfo {author} {\bibfnamefont {G.}~\bibnamefont
  {Strinati}}, \bibinfo {author} {\bibfnamefont {H.~J.}\ \bibnamefont
  {Mattausch}},\ and\ \bibinfo {author} {\bibfnamefont {W.}~\bibnamefont
  {Hanke}},\ }\bibfield  {title} {\bibinfo {title} {Dynamical correlation
  effects on the quasiparticle bloch states of a covalent crystal},\ }\href
  {https://doi.org/10.1103/PhysRevLett.45.290} {\bibfield  {journal} {\bibinfo
  {journal} {Phys. Rev. Lett.}\ }\textbf {\bibinfo {volume} {45}},\ \bibinfo
  {pages} {290} (\bibinfo {year} {1980})}\BibitemShut {NoStop}%
\bibitem [{\citenamefont {Martin}\ \emph {et~al.}(2016)\citenamefont {Martin},
  \citenamefont {Reining},\ and\ \citenamefont
  {Ceperley}}]{martin2016interacting}%
  \BibitemOpen
  \bibfield  {author} {\bibinfo {author} {\bibfnamefont {R.~M.}\ \bibnamefont
  {Martin}}, \bibinfo {author} {\bibfnamefont {L.}~\bibnamefont {Reining}},\
  and\ \bibinfo {author} {\bibfnamefont {D.~M.}\ \bibnamefont {Ceperley}},\
  }\href@noop {} {\emph {\bibinfo {title} {Interacting electrons}}}\ (\bibinfo
  {publisher} {Cambridge University Press},\ \bibinfo {year}
  {2016})\BibitemShut {NoStop}%
\end{thebibliography}%

\newpage

\appendix
\section{}
\label{S}

This Appendix provides additional information on the methodology used in the main text, as it follows. Section \ref{sec:Screening} recalls the main results for the accurate screening model derived in Ref.~\citenum{Sesti2022} and referred to in Fig.~\ref{f_wf}(a) through the `CNT' label. Section \ref{sec:Self} details the self-energy correction $\Sigma(k)$ implemented in the BSE calculation. Section \ref{sec:Exc-Gap} explains the self-consistent calculation of interband coherence, $\Delta(k)$, which is the order parameter of the excitonic insulator phase.

\subsection{Screening model}
\label{sec:Screening}

In this section, we summarize the CNT screening model presented in Ref.~ \citenum{Sesti2022}, used to compute the screened electron-hole interaction, $W$. The main novelty is the inclusion of the local field corrections into the Coulomb potential, which originate from the full three-dimensional topology of the Bloch states, $\Psi_{\alpha\tau k}(\mathbf{r})$. In contrast, in the standard effective mass (EM) treatment the Bloch states are graphene-like, i.e, two dimensional. This we accomplish by working with periodically repeated cylindrical supercells. We fix the total length of the tube along the axial direction as $A= N\, a\, \cos \left(\pi/6-\theta \right)$, with $a= 2.46$ \AA \ being the graphene lattice constant. In the non-periodic directions, the supercells are arranged as a square super-lattice of side $\mathfrak{R}$. In order to prevent any superposition between wave functions of different replicas, we set $\mathfrak{R}=7 R$, where $R$ is the radius of the nanotube. The reciprocal lattice vectors of the resulting three-dimensional lattice are given by:
\begin{align}
\label{eq.G}
\boldsymbol{G}_{\perp} = \frac{\pi}{\mathfrak{R} } \left( n_1 \hat{x} + n_3 \hat{z} \right), \quad \boldsymbol{G}_{\parallel} = \frac{2 \pi}{a \cos \left(\pi/6-\theta \right)} n_2 \hat{y}.
\end{align}
The Bloch states $\Psi$ are expanded over the $\mathbf{G}$ vectors of the  three-dimensional supercell lattice. This allows to compute the Coulomb potential $W(\boldsymbol{r},\boldsymbol{r'})$ of the tube by expanding it over a plane-wave basis set:
\begin{eqnarray}
\label{eq.Wtot}
W(\boldsymbol{r},\boldsymbol{r'} \! ) & = & \sum_{\boldsymbol{q}, \boldsymbol{G}, \boldsymbol{G'}} \! \! \! \! e^{-i (\boldsymbol{G'} + \boldsymbol{q}) \cdot \boldsymbol{r'}} \! e^{i (\boldsymbol{G} + \boldsymbol{q}) \cdot \boldsymbol{r}} \,\epsilon^{-1}_{\boldsymbol{G},\boldsymbol{G'}}(\boldsymbol{q}) \nonumber \\ &&\quad \times\quad  v(\boldsymbol{q} + \boldsymbol{G'}),\nonumber \\ 
\end{eqnarray} 
where $\epsilon^{-1}_{\boldsymbol{G},\boldsymbol{G'}}(\boldsymbol{q})$ is the inverse dielectric function and $v(\boldsymbol{q} + \boldsymbol{G'})$ is the Fourier transform of the bare Coulomb potential. We prevent any spurious Coulomb interaction between replicas by truncating the bare Coulomb potential along the non-periodic directions at the supercells borders, as detailed in Ref.~\cite{rozzi2006exact}. The Fourier transform of the truncated Coulomb potential is:
\begin{multline}
\label{Coulomb}
v(\mathbf{q} + \mathbf{G}) = \frac{4 e^2}{A \mathfrak{R}^2 \left|\mathbf{q} + \mathbf{G}\right|^2}\Big[1\,+\,\\ \mathfrak{R}\, G_{\perp}  J_1(\mathfrak{R} G_{\perp}) \,K_0(\mathfrak{R} |q    + G_{\parallel}|)\\
- \mathfrak{R} \left|q + G_{\parallel}\right| J_0(\mathfrak{R} G_{\perp}) K_1(\mathfrak{R} |q + G_{\parallel}|)\Big],
\end{multline}
where $J_0(x), J_1(x)$ are the Bessel functions of first kind, and $K_0(x), K_1(x)$ the modified Bessel function of second kind. As the Fourier component \eqref{Coulomb} decreases with the magnitude of the reciprocal lattice vector, the Coulomb potential \eqref{eq.Wtot} converges with a limited number of $\mathbf{G}$ vectors. We include just the first $G_{\parallel}$ vector $|n_2| \leq 1 $ and a number of $G_{\perp}$ vectors between $-15 \leq n_1, n_3 \leq 15$.

The inverse dielectric function occuring in \eqref{eq.Wtot} is evaluated within the random phase approximation (RPA):\begin{eqnarray}
\label{RPA_1}
\epsilon_{\boldsymbol{G},\boldsymbol{G'}}(\boldsymbol{q}) = \delta_{\boldsymbol{G},\boldsymbol{G'}} - \Pi_{\boldsymbol{G},\boldsymbol{G'}}(q) \,v(\boldsymbol{q} + \boldsymbol{G}),
\end{eqnarray} 
with polarization:
\begin{multline}\label{eq:pol_corr}
\Pi_{\boldsymbol{G},\boldsymbol{G'}}(q) \quad = \quad  \Pi^{\text{NT}}_{\boldsymbol{G},\boldsymbol{G'}}(q)  \\ \times\quad \left\{5 \cos \! \left[2.7\left(\pi/6- \theta\right) \! \right] \! R q + 3.806\, [R/(\text{1 nm})]^{1.46}  \right\}.
\end{multline} 
Here, the model expression for the polarization 
\begin{multline}
\Pi^{\text{NT}}_{\boldsymbol{G},\boldsymbol{G'}}(q) =  -\frac{2}{\pi \gamma} J_0(R G_{\perp}) J_0(R G'_{\perp}) \\ \sum_{\tau} \Bigg[ 1 +   \frac{ 2 k_{\tau}^2}{q\sqrt{q^2 + 4 k_{\tau}^2}} \log \Bigg(\frac{\sqrt{q^2 + 4 k_{\tau}^2} - q}{\sqrt{q^2 + 4 k_{\tau}^2} + q}\Bigg) \Bigg]
\end{multline}  
is modified according to \eqref{eq:pol_corr}, with the inclusion of a $q-$ and $\theta$-dependent multiplicative correction factor that slighlty differs from unity. This multiplicative factor has been determined fitting first-principles results, specifically for the $G$ components of $\epsilon^{-1}_{\boldsymbol{G},\boldsymbol{G'}}(q)$ (details can be found in Ref.~\citenum{Sesti2022}).

The interaction matrix element associated with electron-hole binding is calculated by projecting the screened potential, $W(\boldsymbol{r},\boldsymbol{r'})$, from an initial $e$-$h$ state $(c, \tau, k)(v, \tau, k +q)$ to a final state $(v,\tau, k)(c,\tau,  k + q)$ within the same valley $\tau$: 
\begin{align}
\label{eq:Wtcnt}
\notag W^{\tau}(k,k+q)= \sum_{\boldsymbol{G}} \sum_{\boldsymbol{G'}} \langle c \tau   k|  e^{-i (\boldsymbol{G'} + \boldsymbol{q}) \cdot \boldsymbol{r'}} | c \tau k+q  \rangle \\ \langle v \tau k+q |    e^{i (\boldsymbol{G} + \boldsymbol{q}) \cdot \boldsymbol{r}} | v \tau k \rangle \ \epsilon^{-1}_{\boldsymbol{G},\boldsymbol{G'}}(q) v(\boldsymbol{q} + \boldsymbol{G'}).
\end{align}
Here the Dirac bra's notation $\left|\alpha \tau k\right>$ stands for the Bloch state $\Psi_{\alpha\tau k}(\mathbf{r})$.
This matrix element can be recasted in a simpler form as: 
\begin{multline}
W^{\tau}(k,k+q) = (\boldsymbol{F}^{\tau \dagger}_{v k} \boldsymbol{F}^{\tau}_{v k+q})   (\boldsymbol{F}^{\tau \dagger }_{c k+q} \boldsymbol{F}^{\tau}_{c k}) \,W(q),
\end{multline}
where 
$W(q)$ is the effective screened Coulomb potential on the tube surface:
\begin{align}
\! \! W(q) \! =  \sum_{\boldsymbol{G}, \boldsymbol{G'}} \! \! J_0(R G_{\perp}) J_0(R G'_{\perp}) \,\epsilon^{-1}_{\boldsymbol{G},\boldsymbol{G'}}(q) \,v(\boldsymbol{q} + \boldsymbol{G'}). 
\end{align}

In the BSE~\eqref{eq.bse}, we include also contributions coming from
short-range scattering processes. In the present case of excitons of spin one, only the short-range intervalley term, $W^{\tau\tau'}$,  is included \cite{Ando2006}. This term is much smaller than $W$ \cite{Ando2006,varsano2017carbon} and can be modeled as a constant in reciprocal space,
\begin{align}
W^{\tau \tau'}(k,k+q) = \frac{\Omega_0 w_2}{4 \pi R},
\end{align}
where $\Omega_0 = (\sqrt{3}/2) a^2$  is the area of graphene unit cell, and the characteristic energy $w_2= 2.6$ eV is estimated from first principles \cite{varsano2017carbon}.

\subsection{Self-energy correction}
\label{sec:Self}

We calculate the self-energy correction for quasiparticles as a first-order perturbation to the electronic states, starting from the screened Hartree-Fock approximation,
\begin{align}
\label{eq.Sigmaa}
\Sigma_{\alpha \tau }(k)=B\! \int\!\!\!\int\!\! d\mathbf{r}\,d\mathbf{r'}
\Psi_{\alpha \tau k }(\boldsymbol{r})\,
 \Sigma^{HF}(\boldsymbol{r},\boldsymbol{r}') \, \Psi^*_{\alpha \tau k}(\boldsymbol{r'}) ,
\end{align}
where we included a rescaling factor $B$ to match results from first principles.
The value $B=1/3$ provides the optimal matching with the quasiparticle energies determined by first-principles GW calculations in \citenum{island2018}.
In the screened Hartree-Fock approximation, the self energy is \cite{strinati1980dynamical,martin2016interacting}:
\begin{align}
\label{eq.HF}
\Sigma^{HF}(\boldsymbol{r},\boldsymbol{r}')=  \lim_{\eta \rightarrow 0^{+}} \frac{i}{2 \pi} \int \! \! d \omega' G_{0}(\boldsymbol{r},\boldsymbol{r}',\omega') W(\boldsymbol{r},\boldsymbol{r}') e^{i \eta \omega'},
\end{align}
where $G_{0}$ is the single-particle propagator, $\omega$ the frequency, and $\eta$ a small positive infinitesimal to ensure the correct time ordering. $G_{0}$ can be written in terms of the single-particle wave functions,
\begin{align}
\label{eq.prop}
G_{0}(\boldsymbol{r},\boldsymbol{r}',\omega) =  \! \! \sum_{\alpha,\tau, k} \!  \frac{\Psi^{*}_{\alpha \tau k }(\boldsymbol{r}) \Psi_{\alpha \tau k}(\boldsymbol{r'})}{\omega-\varepsilon_{\alpha}(k) \! + i \eta \ \textrm{sign}[\varepsilon_{\alpha}(k)]},
\end{align}
with $\varepsilon_{\alpha}(k)$ being the energy of the Bloch state.

Replacing the expression for $\Sigma^{HF}$ and $G_{0}$ in Eq.~\eqref{eq.Sigmaa}, one finds:
\begin{align}
\Sigma_{\alpha}(k)= - \frac{1}{
3 A} \sum_q  | \boldsymbol{F}^{\tau \dagger}_{\alpha k} \boldsymbol{F}^{\tau}_{v k+q}|^2  W(q),
\end{align}
where we dropped the $\tau$ index, as $\Sigma_{\alpha}(k)$ does not depend on the valley. The self-energy of the $e$-$h$ excitation is 
\begin{multline}
\Sigma(k)= \Sigma_{c}(k)-\Sigma_{v}(k) = \\ \frac{1}{3 A} \sum_q (| \boldsymbol{F}^{\tau \dagger}_{v k} \boldsymbol{F}^{\tau}_{v k+q}|^2-| \boldsymbol{F}^{\tau \dagger}_{c k} \boldsymbol{F}^{\tau}_{v k+q}  |^2 )\, W(q).
\end{multline}

\subsection{Interband coherence in the excitonic insulator phase}
\label{sec:Exc-Gap}

The ground-state wavefunction of the excitonic insulator, $|\Psi \rangle$, is similar to the one of the BCS superconductor, provided one replaces the Cooper pairs with the $e$-$h$ pairs,  
\begin{align}
|\Psi \rangle = \prod_{\tau k \sigma} \left[u_{\tau k } + {\text{e}}^{i\phi} v_{\tau k}\sum_{\ell\sigma'}[n_{\ell}\sigma_{\ell}]_{\sigma,\sigma'}  \,\hat{c}^{\tau \dagger}_{k \sigma'} \hat{v}^{\tau}_{k \sigma} \right] \left| 0 \right>.
\end{align}
The pairs making the condensate, $\hat{c}^{\tau \dagger}_{k \sigma'} \hat{v}^{\tau}_{k \sigma}  | 0 \rangle$, are excited from the band insulator ground state,  $| 0 \rangle$, by transferring an electron from a filled Bloch state of momentum $k$, valley $\tau$, spin $\sigma$ to an empty state of spin $\sigma'$ and like momentum and valley (here $\hat{v}$ and $\hat{c}$ are the corresponding Fermi annihilation operators). The positive coherence factors, $u_{\tau k}>0$ and $v_{\tau k}>0$, are the probability amplitudes for finding a valence state of momentum $k$ and valley $\tau$ respectively filled and empty, with $u_{\tau k}^2+v_{\tau k}^2=1$, irrespective of spin. Here $\sigma_{\ell}$ is the 2 $\times$ 2 Pauli matrix along $\ell =x,y,z$, the unit vector $(n_x,n_y,n_z)$ is the direction of the arbitrary spin polarization of the spin-triplet excitons, and $\phi$ is the arbitrary phase of the interband coherence.

The Hamiltonian $\mathcal{H}$ of carbon nanotubes includes the one-body crystal potential term and the two-body Coulomb interaction. The latter is the sum of the long-range screened term and short-range inter-valley interaction:
\begin{multline}
\label{eq.ham}
\mathcal{H}=\sum_{\tau k \sigma} E^{0}_{\tau k} \left( c^{\tau \dagger}_{k \sigma} c^{\tau}_{k \sigma}- v^{\tau \dagger}_{k \sigma} v^{\tau}_{k \sigma} \right) \\ -  \frac{1}{A} \sum_{\tau k k'} \sum_{\sigma \sigma'} W^{\tau}(k,k') \ \left( c^{\tau \dagger}_{k \sigma}  v^{\tau}_{k \sigma'} v^{\tau \dagger}_{k' \sigma'} c^{\tau}_{k' \sigma} + c.c. \right) \\ - \frac{1}{A} \sum_{\tau \neq \tau'}  \sum_{k k'} \sum_{\sigma \sigma'} W^{\tau \tau'} \left( c^{\tau \dagger}_{k \sigma'} v^{\tau}_{k \sigma}  v^{\tau' \dagger}_{k' \sigma'} c^{\tau' }_{k' \sigma} + c.c. \right).
\end{multline}
Here $E^{0}_{\tau k}=\gamma \left[k_c^2+k^2\right]^{1/2} 
+  \Sigma(k)/2$ is the energy of conduction-band quasiparticles. Whereas the sum over $k$ and $k'$ in principle extends throughout the Brillouin zone, practically
one may limit it to $-0.1\times2\pi/a < k < 0.1\times2\pi/a $, due to the large Fermi velocity of the Dirac cone. 

The energy of the non-interacting band insulator, $\mathcal{E}_{0}$, is:
\begin{align}
 \mathcal{E}_{0}= -2 \sum_{\tau k} E^{0}_{\tau k}.
\end{align}
The energy of the EI state, $\mathcal{E}_{\text{EI}}$, is lower than $\mathcal{E}_{0}$ by the amount
\begin{multline}
\mathcal{E}_{\text{EI}}-\mathcal{E}_{0}= 2 \sum_{\tau k} E^{0}_{\tau k}  (v_{\tau k}^2- u_{\tau k}^2+1) \\ - \frac{4}{A} \sum_{\tau k k'} W^{\tau}(k,k')  v_{\tau k} u_{\tau k}  v_{\tau k'} u_{\tau k'}   \\ - \frac{4}{A} \sum_{\tau \neq \tau'} \sum_{k k'}  W^{\tau \tau'} v_{\tau k } u_{\tau k} v_{\tau' k'} u_{\tau' k'}. 
\end{multline}
We introduce the interband coherence, $\Delta(\tau k)$, as 
\begin{align}
\label{eq.delta}
\Delta(\tau k )  =  \frac{ {\text{e}}^{i\phi} }{A} \sum_{k'}  \left[   W^{\tau}(k,k')  v_{\tau k'} u_{\tau k'} + \sum_{\tau'\neq \tau} W^{\tau \tau'} \! v_{\tau' k'} u_{\tau' k'}   \right]   ,
\end{align}
through which we can rewrite the EI energy gain as
\begin{align}
\mathcal{E}_{\text{EI}}-\mathcal{E}_{0}=  2 \sum_{\tau k} \left[ E^{0}_{\tau k} ( v_{\tau k}^2 -u_{\tau k}^2 +1)  - 2 \left|\Delta(\tau k)\right| u_{\tau k} v_{\tau k} \right]. 
\end{align}
The coherence factors $u_{\tau k}$ and $v_{\tau k}$ are variational quantities that are determined through the Lagrange multiplier method, by means of minimizing the energy gain $\mathcal{E}_{\text{EI}}-\mathcal{E}_{0}$,
\begin{align}
u_{\tau k}^2=& \ \frac{1}{2} \left( 1+\frac{E^{0}_{\tau k}}{2 E_{\tau k}} \right),
\end{align}
with 
\begin{equation}
E_{\tau k} =\left[(E^{0}_{\tau k})^2 + \left|\Delta(\tau k)\right|^2\right]^{1/2}.
\end{equation}
The knowledge of coherence factors allows us to rewrite \eqref{eq.delta} as the self-consistent gap equation,
\begin{align}
\label{eq.selfcons}
\Delta(\tau k)= \frac{1}{A} \sum_{k'} \left[  \frac{ W^{\tau}(k,k') \Delta(\tau k')}{E_{\tau k'}}   + \sum_{\tau' \neq \tau} W^{\tau \tau'} \frac{\Delta(\tau' k')}{E_{\tau' k'}}  \right].
\end{align}
Note the similarity between \eqref{eq.selfcons} and the BSE \eqref{eq.bse}, which becomes apparent after introducing a pseudo wave function, $\varphi_{\tau}(k)$, as
\begin{eqnarray}
\varphi_{\tau}(k)= \frac{\Delta(\tau k)}{2 E_{\tau k} },
\end{eqnarray}
which leads to 
\begin{multline}
 E_{\tau k}\, \varphi_{\tau}(k) - \frac{1}{A} \sum_{k'} W^{\tau}(k,k') \varphi_{\tau}(k') \\
-  \frac{1}{A} \sum_{\tau' \neq \tau} \sum_{k'} W^{\tau \tau'} \varphi_{\tau'}(k')  =0.
\end{multline}
The latter equation is solved numerically, starting from a seed for $\Delta(\tau k)$, updated until convergence is met.  We make use of the eigenstates and eigenvalues of the Bethe-Salpeter equation for the seed:
\begin{eqnarray}\label{eq:delta_start}
\Delta^{\text{seed}}(\tau k)= \frac{E^{0}_{\tau k} - \mathcal{E}_u}{2} \left|\frac{\psi_{\tau}(k)}{\psi_{\tau}(0)} \right|.
\end{eqnarray}

\end{document}